\documentclass{llncs}

\usepackage[latin1]{inputenc}

\usepackage[english]{babel}

% ...

\usepackage{color}

% ...

\usepackage{cite}

\usepackage[final]{microtype}

\usepackage{changebar}

\usepackage{etoolbox}

\usepackage{fixltx2e}

\usepackage{xargs}

\usepackage{xspace}

\usepackage{xstring}

\usepackage[cmex10]{mathtools}
\usepackage{amssymb}
\usepackage{amsfonts}
\usepackage{mathrsfs}
\usepackage{latexsym}
\usepackage{textcomp}
\usepackage{pifont}

\newcommand{\argemp}[2]
	{\if&#1&\else#2\fi}

\newcommand{\argdef}[2]
	{\if&#1&#2\else#1\fi}

\newcommand{\argint}[3]
	{\if&#2&\else#1#2#3\fi}

\newcommand{\argext}[3]
	{\if&#1&#3\else#1\if&#3&\else#2#3\fi\fi}

\newcommandx{\argsubsup}[3][2=, 3=]
	{\def\argsubscript{{#2}}\def\argsuperscript{{#3}}#1}

\newcommandx{\argind}[9][2=, 3=, 4=, 5=, 6=, 7=, 8=, 9=]
	{%
	\switch[#1=]%
		\case{0}#2%
		\case{1}#3%
		\case{2}#4%
		\case{3}#5%
		\case{4}#6%
		\case{5}#7%
		\case{6}#8%
		\case{7}#9%
		\otherwise\ensuremath{\clubsuit}%
	\endswitch%
	}

\newcommand{\arga}[1]
	{#1}
\newcommand{\argb}[2]
	{\argext{\arga{#1}}{, \allowbreak}{#2}}
\newcommand{\argc}[3]
	{\argext{\argb{#1}{#2}}{, \allowbreak}{#3}}
\newcommand{\argd}[4]
	{\argext{\argc{#1}{#2}{#3}}{, \allowbreak}{#4}}
\newcommand{\arge}[5]
	{\argext{\argd{#1}{#2}{#3}{#4}}{, \allowbreak}{#5}}
\newcommand{\argf}[6]
	{\argext{\arge{#1}{#2}{#3}{#4}{#5}}{, \allowbreak}{#6}}
\newcommand{\argg}[7]
	{\argext{\argf{#1}{#2}{#3}{#4}{#5}{#6}}{, \allowbreak}{#7}}
\newcommand{\argh}[8]
	{\argext{\argg{#1}{#2}{#3}{#4}{#5}{#6}{#7}}{, \allowbreak}{#8}}
\newcommand{\argi}[9]
	{\argext{\argh{#1}{#2}{#3}{#4}{#5}{#6}{#7}{#8}}{, \allowbreak}{#9}}

\newcommand{\txtfnt}[2][]
	{{%
	\IfStrEq{#1}{}
		{#2}
		{%
		\StrLeft{#1}{2}[\optbgn]%
		\StrGobbleLeft{#1}{2}[\optend]%
		\IfStrEqCase{\optbgn}
			{%
			{Rm}{\rmfamily\txtfnt[\optend]{#2}}%
			{Sf}{\sffamily\txtfnt[\optend]{#2}}%
			{Tt}{\ttfamily\txtfnt[\optend]{#2}}%
			{Up}{\upshape\txtfnt[\optend]{#2}}%
			{It}{\itshape\txtfnt[\optend]{#2}}%
			{Sl}{\slshape\txtfnt[\optend]{#2}}%
			{Sc}{\scshape\txtfnt[\optend]{#2}}%
			{Md}{\mdseries\txtfnt[\optend]{#2}}%
			{Bf}{\bfseries\txtfnt[\optend]{#2}}%
			{Em}{\emph{\txtfnt[\optend]{#2}}}%
			}
			[\ensuremath{\clubsuit}]%
		}%
	}}

\newcommand{\txtsub}[2][]
	{\argemp{#2}{\ensuremath{_{\text{\txtfnt[#1]{#2}}}}}}

\newcommand{\txtsup}[2][]
	{\argemp{#2}{\ensuremath{^{\text{\txtfnt[#1]{#2}}}}}}

\newcommandx{\txt}[4][1=, 3=, 4=]
	{%
	\ensuremath{\text{%
		\txtfnt[#1]{#2}\ensuremath{\txtsub[#1]{#3}\txtsup[#1]{#4}}%
	}}%
	}

\newcommandx{\txtarg}[5][1=, 3=, 4=]
	{{\txt[#1]{#2}[#3][#4]\argint{(}{#5}{)}}}

\newcommand{\txtstyname}{RmScMd}
\newcommand{\txtname}[1][]
	{\txt[\argdef{#1}{\txtstyname}]}
\newcommand{\txtargname}[1][]
	{\txtarg[\argdef{#1}{\txtstyname}]}

\newcommand{\txtstyabr}{Em}
\newcommand{\txtabr}[1][]
	{\txt[\argdef{#1}{\txtstyabr}]}

\newcommandx{\mthfnt}[3][1=, 2=0]
	{{%
	\IfStrEqCase{#1}
		{%
		{}%
			{#3}%
		{Name}%
			{%
			\IfStrEqCase{#2}
				{%
				{0}{\mathcal{#3}}%
				{1}{\mathscr{#3}}%
				{2}{\mathfrak{#3}}%
				{3}{\mathbb{#3}}%
				}
				[\ensuremath{\clubsuit}]%
			}%
		{Set}%
			{%
			\IfStrEqCase{#2}
				{%
				{0}{\mathrm{#3}}%
				{1}{\mathsf{#3}}%
				{2}{\mathbb{#3}}%
				{3}{\mathbf{#3}}%
				}
				[\ensuremath{\clubsuit}]%
			}%
		{Fun}%
			{%
			\IfStrEqCase{#2}
				{%
				{0}{\mathsf{#3}}%
				{1}{\mathrm{#3}}%
				}
				[\ensuremath{\clubsuit}]%
			}%
		{Rel}%
			{%
			\IfStrEqCase{#2}
				{%
				{0}{\mathit{#3}}%
				{1}{\mathtt{#3}}%
				}
				[\ensuremath{\clubsuit}]%
			}%
		{Sym}%
			{%
			\IfStrEqCase{#2}
				{%
				{0}{\mathtt{#3}}%
				{1}{\mathbf{#3}}%
				}
				[\ensuremath{\clubsuit}]%
			}%
		{Elm}%
			{\mathnormal{#3}}
		}
		[\ensuremath{\clubsuit}]%
	}}

\newcommand{\mthsub}[1]
	{\argemp{#1}{\ensuremath{_{\mathnormal{#1}}}}}

\newcommand{\mthsup}[1]
	{\argemp{#1}{\ensuremath{^{\mathnormal{#1}}}}}

\newcommandx{\mth}[5][1=, 2=0, 4=, 5=]
	{{\ensuremath{\mthfnt[#1][#2]{#3}\mthsub{#4}\mthsup{#5}}}}

\newcommandx{\mtharg}[6][1=, 2=0, 4=, 5=]
	{{\mth[#1][#2]{#3}[#4][#5]\ensuremath{\argint{(}{#6}{)}}}}

\newcommand{\mthempty}
	{\mth[][]}

\newcommand{\mthstyname}{0}
\newcommand{\mthname}[1][]
	{\mth[Name][\argdef{#1}{\mthstyname}]}

\newcommand{\mthstyset}{0}
\newcommand{\mthset}[1][]
	{\mth[Set][\argdef{#1}{\mthstyset}]}
\newcommand{\mthargset}[1][]
	{\mtharg[Set][\argdef{#1}{\mthstyset}]}

\newcommand{\mthstyfun}{0}
\newcommand{\mthfun}[1][]
	{\mth[Fun][\argdef{#1}{\mthstyfun}]}
\newcommand{\mthargfun}[1][]
	{\mtharg[Fun][\argdef{#1}{\mthstyfun}]}

\newcommand{\mthstyrel}{0}
\newcommand{\mthrel}[1][]
	{\mth[Rel][\argdef{#1}{\mthstyrel}]}

\newcommand{\mthstysym}{0}
\newcommand{\mthsym}[1][]
	{\mth[Sym][\argdef{#1}{\mthstysym}]}

\newcommand{\mthstyelm}{0}
\newcommand{\mthelm}[1][]
	{\mth[Elm][\argdef{#1}{\mthstyelm}]}

\newcommandx{\AName}[4][1=, 2=, 3=, 4=]{\mthname[#4]{A#3}[#1][#2]}
\newcommandx{\BName}[4][1=, 2=, 3=, 4=]{\mthname[#4]{B#3}[#1][#2]}
\newcommandx{\CName}[4][1=, 2=, 3=, 4=]{\mthname[#4]{C#3}[#1][#2]}
\newcommandx{\DName}[4][1=, 2=, 3=, 4=]{\mthname[#4]{D#3}[#1][#2]}
\newcommandx{\EName}[4][1=, 2=, 3=, 4=]{\mthname[#4]{E#3}[#1][#2]}
\newcommandx{\FName}[4][1=, 2=, 3=, 4=]{\mthname[#4]{F#3}[#1][#2]}
\newcommandx{\GName}[4][1=, 2=, 3=, 4=]{\mthname[#4]{G#3}[#1][#2]}
\newcommandx{\HName}[4][1=, 2=, 3=, 4=]{\mthname[#4]{H#3}[#1][#2]}
\newcommandx{\IName}[4][1=, 2=, 3=, 4=]{\mthname[#4]{I#3}[#1][#2]}
\newcommandx{\JName}[4][1=, 2=, 3=, 4=]{\mthname[#4]{J#3}[#1][#2]}
\newcommandx{\KName}[4][1=, 2=, 3=, 4=]{\mthname[#4]{K#3}[#1][#2]}
\newcommandx{\LName}[4][1=, 2=, 3=, 4=]{\mthname[#4]{L#3}[#1][#2]}
\newcommandx{\MName}[4][1=, 2=, 3=, 4=]{\mthname[#4]{M#3}[#1][#2]}
\newcommandx{\NName}[4][1=, 2=, 3=, 4=]{\mthname[#4]{N#3}[#1][#2]}
\newcommandx{\OName}[4][1=, 2=, 3=, 4=]{\mthname[#4]{O#3}[#1][#2]}
\newcommandx{\PName}[4][1=, 2=, 3=, 4=]{\mthname[#4]{P#3}[#1][#2]}
\newcommandx{\QName}[4][1=, 2=, 3=, 4=]{\mthname[#4]{Q#3}[#1][#2]}
\newcommandx{\RName}[4][1=, 2=, 3=, 4=]{\mthname[#4]{R#3}[#1][#2]}
\newcommandx{\SName}[4][1=, 2=, 3=, 4=]{\mthname[#4]{S#3}[#1][#2]}
\newcommandx{\TName}[4][1=, 2=, 3=, 4=]{\mthname[#4]{T#3}[#1][#2]}
\newcommandx{\UName}[4][1=, 2=, 3=, 4=]{\mthname[#4]{U#3}[#1][#2]}
\newcommandx{\VName}[4][1=, 2=, 3=, 4=]{\mthname[#4]{V#3}[#1][#2]}
\newcommandx{\WName}[4][1=, 2=, 3=, 4=]{\mthname[#4]{W#3}[#1][#2]}
\newcommandx{\XName}[4][1=, 2=, 3=, 4=]{\mthname[#4]{X#3}[#1][#2]}
\newcommandx{\YName}[4][1=, 2=, 3=, 4=]{\mthname[#4]{Y#3}[#1][#2]}
\newcommandx{\ZName}[4][1=, 2=, 3=, 4=]{\mthname[#4]{Z#3}[#1][#2]}

\newcommandx{\ASet}[4][1=, 2=, 3=, 4=]{\mthset[#4]{A#3}[#1][#2]}
\newcommandx{\BSet}[4][1=, 2=, 3=, 4=]{\mthset[#4]{B#3}[#1][#2]}
\newcommandx{\CSet}[4][1=, 2=, 3=, 4=]{\mthset[#4]{C#3}[#1][#2]}
\newcommandx{\DSet}[4][1=, 2=, 3=, 4=]{\mthset[#4]{D#3}[#1][#2]}
\newcommandx{\ESet}[4][1=, 2=, 3=, 4=]{\mthset[#4]{E#3}[#1][#2]}
\newcommandx{\FSet}[4][1=, 2=, 3=, 4=]{\mthset[#4]{F#3}[#1][#2]}
\newcommandx{\GSet}[4][1=, 2=, 3=, 4=]{\mthset[#4]{G#3}[#1][#2]}
\newcommandx{\HSet}[4][1=, 2=, 3=, 4=]{\mthset[#4]{H#3}[#1][#2]}
\newcommandx{\ISet}[4][1=, 2=, 3=, 4=]{\mthset[#4]{I#3}[#1][#2]}
\newcommandx{\JSet}[4][1=, 2=, 3=, 4=]{\mthset[#4]{J#3}[#1][#2]}
\newcommandx{\KSet}[4][1=, 2=, 3=, 4=]{\mthset[#4]{K#3}[#1][#2]}
\newcommandx{\LSet}[4][1=, 2=, 3=, 4=]{\mthset[#4]{L#3}[#1][#2]}
\newcommandx{\MSet}[4][1=, 2=, 3=, 4=]{\mthset[#4]{M#3}[#1][#2]}
\newcommandx{\NSet}[4][1=, 2=, 3=, 4=]{\mthset[#4]{N#3}[#1][#2]}
\newcommandx{\OSet}[4][1=, 2=, 3=, 4=]{\mthset[#4]{O#3}[#1][#2]}
\newcommandx{\PSet}[4][1=, 2=, 3=, 4=]{\mthset[#4]{P#3}[#1][#2]}
\newcommandx{\QSet}[4][1=, 2=, 3=, 4=]{\mthset[#4]{Q#3}[#1][#2]}
\newcommandx{\RSet}[4][1=, 2=, 3=, 4=]{\mthset[#4]{R#3}[#1][#2]}
\newcommandx{\SSet}[4][1=, 2=, 3=, 4=]{\mthset[#4]{S#3}[#1][#2]}
\newcommandx{\TSet}[4][1=, 2=, 3=, 4=]{\mthset[#4]{T#3}[#1][#2]}
\newcommandx{\USet}[4][1=, 2=, 3=, 4=]{\mthset[#4]{U#3}[#1][#2]}
\newcommandx{\VSet}[4][1=, 2=, 3=, 4=]{\mthset[#4]{V#3}[#1][#2]}
\newcommandx{\WSet}[4][1=, 2=, 3=, 4=]{\mthset[#4]{W#3}[#1][#2]}
\newcommandx{\XSet}[4][1=, 2=, 3=, 4=]{\mthset[#4]{X#3}[#1][#2]}
\newcommandx{\YSet}[4][1=, 2=, 3=, 4=]{\mthset[#4]{Y#3}[#1][#2]}
\newcommandx{\ZSet}[4][1=, 2=, 3=, 4=]{\mthset[#4]{Z#3}[#1][#2]}

\newcommandx{\aSet}[4][1=, 2=, 3=, 4=]{\mthset[#4]{a#3}[#1][#2]}
\newcommandx{\bSet}[4][1=, 2=, 3=, 4=]{\mthset[#4]{b#3}[#1][#2]}
\newcommandx{\cSet}[4][1=, 2=, 3=, 4=]{\mthset[#4]{c#3}[#1][#2]}
\newcommandx{\dSet}[4][1=, 2=, 3=, 4=]{\mthset[#4]{d#3}[#1][#2]}
\newcommandx{\eSet}[4][1=, 2=, 3=, 4=]{\mthset[#4]{e#3}[#1][#2]}
\newcommandx{\fSet}[4][1=, 2=, 3=, 4=]{\mthset[#4]{f#3}[#1][#2]}
\newcommandx{\gSet}[4][1=, 2=, 3=, 4=]{\mthset[#4]{g#3}[#1][#2]}
\newcommandx{\hSet}[4][1=, 2=, 3=, 4=]{\mthset[#4]{h#3}[#1][#2]}
\newcommandx{\iSet}[4][1=, 2=, 3=, 4=]{\mthset[#4]{i#3}[#1][#2]}
\newcommandx{\jSet}[4][1=, 2=, 3=, 4=]{\mthset[#4]{j#3}[#1][#2]}
\newcommandx{\kSet}[4][1=, 2=, 3=, 4=]{\mthset[#4]{k#3}[#1][#2]}
\newcommandx{\lSet}[4][1=, 2=, 3=, 4=]{\mthset[#4]{l#3}[#1][#2]}
\newcommandx{\mSet}[4][1=, 2=, 3=, 4=]{\mthset[#4]{m#3}[#1][#2]}
\newcommandx{\nSet}[4][1=, 2=, 3=, 4=]{\mthset[#4]{n#3}[#1][#2]}
\newcommandx{\oSet}[4][1=, 2=, 3=, 4=]{\mthset[#4]{o#3}[#1][#2]}
\newcommandx{\pSet}[4][1=, 2=, 3=, 4=]{\mthset[#4]{p#3}[#1][#2]}
\newcommandx{\qSet}[4][1=, 2=, 3=, 4=]{\mthset[#4]{q#3}[#1][#2]}
\newcommandx{\rSet}[4][1=, 2=, 3=, 4=]{\mthset[#4]{r#3}[#1][#2]}
\newcommandx{\sSet}[4][1=, 2=, 3=, 4=]{\mthset[#4]{s#3}[#1][#2]}
\newcommandx{\tSet}[4][1=, 2=, 3=, 4=]{\mthset[#4]{t#3}[#1][#2]}
\newcommandx{\uSet}[4][1=, 2=, 3=, 4=]{\mthset[#4]{u#3}[#1][#2]}
\newcommandx{\vSet}[4][1=, 2=, 3=, 4=]{\mthset[#4]{v#3}[#1][#2]}
\newcommandx{\wSet}[4][1=, 2=, 3=, 4=]{\mthset[#4]{w#3}[#1][#2]}
\newcommandx{\xSet}[4][1=, 2=, 3=, 4=]{\mthset[#4]{x#3}[#1][#2]}
\newcommandx{\ySet}[4][1=, 2=, 3=, 4=]{\mthset[#4]{y#3}[#1][#2]}
\newcommandx{\zSet}[4][1=, 2=, 3=, 4=]{\mthset[#4]{z#3}[#1][#2]}

\newcommandx{\AFun}[4][1=, 2=, 3=, 4=]{\mthfun[#4]{A#3}[#1][#2]}
\newcommandx{\BFun}[4][1=, 2=, 3=, 4=]{\mthfun[#4]{B#3}[#1][#2]}
\newcommandx{\CFun}[4][1=, 2=, 3=, 4=]{\mthfun[#4]{C#3}[#1][#2]}
\newcommandx{\DFun}[4][1=, 2=, 3=, 4=]{\mthfun[#4]{D#3}[#1][#2]}
\newcommandx{\EFun}[4][1=, 2=, 3=, 4=]{\mthfun[#4]{E#3}[#1][#2]}
\newcommandx{\FFun}[4][1=, 2=, 3=, 4=]{\mthfun[#4]{F#3}[#1][#2]}
\newcommandx{\GFun}[4][1=, 2=, 3=, 4=]{\mthfun[#4]{G#3}[#1][#2]}
\newcommandx{\HFun}[4][1=, 2=, 3=, 4=]{\mthfun[#4]{H#3}[#1][#2]}
\newcommandx{\IFun}[4][1=, 2=, 3=, 4=]{\mthfun[#4]{I#3}[#1][#2]}
\newcommandx{\JFun}[4][1=, 2=, 3=, 4=]{\mthfun[#4]{J#3}[#1][#2]}
\newcommandx{\KFun}[4][1=, 2=, 3=, 4=]{\mthfun[#4]{K#3}[#1][#2]}
\newcommandx{\LFun}[4][1=, 2=, 3=, 4=]{\mthfun[#4]{L#3}[#1][#2]}
\newcommandx{\MFun}[4][1=, 2=, 3=, 4=]{\mthfun[#4]{M#3}[#1][#2]}
\newcommandx{\NFun}[4][1=, 2=, 3=, 4=]{\mthfun[#4]{N#3}[#1][#2]}
\newcommandx{\OFun}[4][1=, 2=, 3=, 4=]{\mthfun[#4]{O#3}[#1][#2]}
\newcommandx{\PFun}[4][1=, 2=, 3=, 4=]{\mthfun[#4]{P#3}[#1][#2]}
\newcommandx{\QFun}[4][1=, 2=, 3=, 4=]{\mthfun[#4]{Q#3}[#1][#2]}
\newcommandx{\RFun}[4][1=, 2=, 3=, 4=]{\mthfun[#4]{R#3}[#1][#2]}
\newcommandx{\SFun}[4][1=, 2=, 3=, 4=]{\mthfun[#4]{S#3}[#1][#2]}
\newcommandx{\TFun}[4][1=, 2=, 3=, 4=]{\mthfun[#4]{T#3}[#1][#2]}
\newcommandx{\UFun}[4][1=, 2=, 3=, 4=]{\mthfun[#4]{U#3}[#1][#2]}
\newcommandx{\VFun}[4][1=, 2=, 3=, 4=]{\mthfun[#4]{V#3}[#1][#2]}
\newcommandx{\WFun}[4][1=, 2=, 3=, 4=]{\mthfun[#4]{W#3}[#1][#2]}
\newcommandx{\XFun}[4][1=, 2=, 3=, 4=]{\mthfun[#4]{X#3}[#1][#2]}
\newcommandx{\YFun}[4][1=, 2=, 3=, 4=]{\mthfun[#4]{Y#3}[#1][#2]}
\newcommandx{\ZFun}[4][1=, 2=, 3=, 4=]{\mthfun[#4]{Z#3}[#1][#2]}

\newcommandx{\aFun}[4][1=, 2=, 3=, 4=]{\mthfun[#4]{a#3}[#1][#2]}
\newcommandx{\bFun}[4][1=, 2=, 3=, 4=]{\mthfun[#4]{b#3}[#1][#2]}
\newcommandx{\cFun}[4][1=, 2=, 3=, 4=]{\mthfun[#4]{c#3}[#1][#2]}
\newcommandx{\dFun}[4][1=, 2=, 3=, 4=]{\mthfun[#4]{d#3}[#1][#2]}
\newcommandx{\eFun}[4][1=, 2=, 3=, 4=]{\mthfun[#4]{e#3}[#1][#2]}
\newcommandx{\fFun}[4][1=, 2=, 3=, 4=]{\mthfun[#4]{f#3}[#1][#2]}
\newcommandx{\gFun}[4][1=, 2=, 3=, 4=]{\mthfun[#4]{g#3}[#1][#2]}
\newcommandx{\hFun}[4][1=, 2=, 3=, 4=]{\mthfun[#4]{h#3}[#1][#2]}
\newcommandx{\iFun}[4][1=, 2=, 3=, 4=]{\mthfun[#4]{i#3}[#1][#2]}
\newcommandx{\jFun}[4][1=, 2=, 3=, 4=]{\mthfun[#4]{j#3}[#1][#2]}
\newcommandx{\kFun}[4][1=, 2=, 3=, 4=]{\mthfun[#4]{k#3}[#1][#2]}
\newcommandx{\lFun}[4][1=, 2=, 3=, 4=]{\mthfun[#4]{l#3}[#1][#2]}
\newcommandx{\mFun}[4][1=, 2=, 3=, 4=]{\mthfun[#4]{m#3}[#1][#2]}
\newcommandx{\nFun}[4][1=, 2=, 3=, 4=]{\mthfun[#4]{n#3}[#1][#2]}
\newcommandx{\oFun}[4][1=, 2=, 3=, 4=]{\mthfun[#4]{o#3}[#1][#2]}
\newcommandx{\pFun}[4][1=, 2=, 3=, 4=]{\mthfun[#4]{p#3}[#1][#2]}
\newcommandx{\qFun}[4][1=, 2=, 3=, 4=]{\mthfun[#4]{q#3}[#1][#2]}
\newcommandx{\rFun}[4][1=, 2=, 3=, 4=]{\mthfun[#4]{r#3}[#1][#2]}
\newcommandx{\sFun}[4][1=, 2=, 3=, 4=]{\mthfun[#4]{s#3}[#1][#2]}
\newcommandx{\tFun}[4][1=, 2=, 3=, 4=]{\mthfun[#4]{t#3}[#1][#2]}
\newcommandx{\uFun}[4][1=, 2=, 3=, 4=]{\mthfun[#4]{u#3}[#1][#2]}
\newcommandx{\vFun}[4][1=, 2=, 3=, 4=]{\mthfun[#4]{v#3}[#1][#2]}
\newcommandx{\wFun}[4][1=, 2=, 3=, 4=]{\mthfun[#4]{w#3}[#1][#2]}
\newcommandx{\xFun}[4][1=, 2=, 3=, 4=]{\mthfun[#4]{x#3}[#1][#2]}
\newcommandx{\yFun}[4][1=, 2=, 3=, 4=]{\mthfun[#4]{y#3}[#1][#2]}
\newcommandx{\zFun}[4][1=, 2=, 3=, 4=]{\mthfun[#4]{z#3}[#1][#2]}

\newcommandx{\ARel}[4][1=, 2=, 3=, 4=]{\mthrel[#4]{A#3}[#1][#2]}
\newcommandx{\BRel}[4][1=, 2=, 3=, 4=]{\mthrel[#4]{B#3}[#1][#2]}
\newcommandx{\CRel}[4][1=, 2=, 3=, 4=]{\mthrel[#4]{C#3}[#1][#2]}
\newcommandx{\DRel}[4][1=, 2=, 3=, 4=]{\mthrel[#4]{D#3}[#1][#2]}
\newcommandx{\ERel}[4][1=, 2=, 3=, 4=]{\mthrel[#4]{E#3}[#1][#2]}
\newcommandx{\FRel}[4][1=, 2=, 3=, 4=]{\mthrel[#4]{F#3}[#1][#2]}
\newcommandx{\GRel}[4][1=, 2=, 3=, 4=]{\mthrel[#4]{G#3}[#1][#2]}
\newcommandx{\HRel}[4][1=, 2=, 3=, 4=]{\mthrel[#4]{H#3}[#1][#2]}
\newcommandx{\IRel}[4][1=, 2=, 3=, 4=]{\mthrel[#4]{I#3}[#1][#2]}
\newcommandx{\JRel}[4][1=, 2=, 3=, 4=]{\mthrel[#4]{J#3}[#1][#2]}
\newcommandx{\KRel}[4][1=, 2=, 3=, 4=]{\mthrel[#4]{K#3}[#1][#2]}
\newcommandx{\LRel}[4][1=, 2=, 3=, 4=]{\mthrel[#4]{L#3}[#1][#2]}
\newcommandx{\MRel}[4][1=, 2=, 3=, 4=]{\mthrel[#4]{M#3}[#1][#2]}
\newcommandx{\NRel}[4][1=, 2=, 3=, 4=]{\mthrel[#4]{N#3}[#1][#2]}
\newcommandx{\ORel}[4][1=, 2=, 3=, 4=]{\mthrel[#4]{O#3}[#1][#2]}
\newcommandx{\PRel}[4][1=, 2=, 3=, 4=]{\mthrel[#4]{P#3}[#1][#2]}
\newcommandx{\QRel}[4][1=, 2=, 3=, 4=]{\mthrel[#4]{Q#3}[#1][#2]}
\newcommandx{\RRel}[4][1=, 2=, 3=, 4=]{\mthrel[#4]{R#3}[#1][#2]}
\newcommandx{\SRel}[4][1=, 2=, 3=, 4=]{\mthrel[#4]{S#3}[#1][#2]}
\newcommandx{\TRel}[4][1=, 2=, 3=, 4=]{\mthrel[#4]{T#3}[#1][#2]}
\newcommandx{\URel}[4][1=, 2=, 3=, 4=]{\mthrel[#4]{U#3}[#1][#2]}
\newcommandx{\VRel}[4][1=, 2=, 3=, 4=]{\mthrel[#4]{V#3}[#1][#2]}
\newcommandx{\WRel}[4][1=, 2=, 3=, 4=]{\mthrel[#4]{W#3}[#1][#2]}
\newcommandx{\XRel}[4][1=, 2=, 3=, 4=]{\mthrel[#4]{X#3}[#1][#2]}
\newcommandx{\YRel}[4][1=, 2=, 3=, 4=]{\mthrel[#4]{Y#3}[#1][#2]}
\newcommandx{\ZRel}[4][1=, 2=, 3=, 4=]{\mthrel[#4]{Z#3}[#1][#2]}

\newcommandx{\aRel}[4][1=, 2=, 3=, 4=]{\mthrel[#4]{a#3}[#1][#2]}
\newcommandx{\bRel}[4][1=, 2=, 3=, 4=]{\mthrel[#4]{b#3}[#1][#2]}
\newcommandx{\cRel}[4][1=, 2=, 3=, 4=]{\mthrel[#4]{c#3}[#1][#2]}
\newcommandx{\dRel}[4][1=, 2=, 3=, 4=]{\mthrel[#4]{d#3}[#1][#2]}
\newcommandx{\eRel}[4][1=, 2=, 3=, 4=]{\mthrel[#4]{e#3}[#1][#2]}
\newcommandx{\fRel}[4][1=, 2=, 3=, 4=]{\mthrel[#4]{f#3}[#1][#2]}
\newcommandx{\gRel}[4][1=, 2=, 3=, 4=]{\mthrel[#4]{g#3}[#1][#2]}
\newcommandx{\hRel}[4][1=, 2=, 3=, 4=]{\mthrel[#4]{h#3}[#1][#2]}
\newcommandx{\iRel}[4][1=, 2=, 3=, 4=]{\mthrel[#4]{i#3}[#1][#2]}
\newcommandx{\jRel}[4][1=, 2=, 3=, 4=]{\mthrel[#4]{j#3}[#1][#2]}
\newcommandx{\kRel}[4][1=, 2=, 3=, 4=]{\mthrel[#4]{k#3}[#1][#2]}
\newcommandx{\lRel}[4][1=, 2=, 3=, 4=]{\mthrel[#4]{l#3}[#1][#2]}
\newcommandx{\mRel}[4][1=, 2=, 3=, 4=]{\mthrel[#4]{m#3}[#1][#2]}
\newcommandx{\nRel}[4][1=, 2=, 3=, 4=]{\mthrel[#4]{n#3}[#1][#2]}
\newcommandx{\oRel}[4][1=, 2=, 3=, 4=]{\mthrel[#4]{o#3}[#1][#2]}
\newcommandx{\pRel}[4][1=, 2=, 3=, 4=]{\mthrel[#4]{p#3}[#1][#2]}
\newcommandx{\qRel}[4][1=, 2=, 3=, 4=]{\mthrel[#4]{q#3}[#1][#2]}
\newcommandx{\rRel}[4][1=, 2=, 3=, 4=]{\mthrel[#4]{r#3}[#1][#2]}
\newcommandx{\sRel}[4][1=, 2=, 3=, 4=]{\mthrel[#4]{s#3}[#1][#2]}
\newcommandx{\tRel}[4][1=, 2=, 3=, 4=]{\mthrel[#4]{t#3}[#1][#2]}
\newcommandx{\uRel}[4][1=, 2=, 3=, 4=]{\mthrel[#4]{u#3}[#1][#2]}
\newcommandx{\vRel}[4][1=, 2=, 3=, 4=]{\mthrel[#4]{v#3}[#1][#2]}
\newcommandx{\wRel}[4][1=, 2=, 3=, 4=]{\mthrel[#4]{w#3}[#1][#2]}
\newcommandx{\xRel}[4][1=, 2=, 3=, 4=]{\mthrel[#4]{x#3}[#1][#2]}
\newcommandx{\yRel}[4][1=, 2=, 3=, 4=]{\mthrel[#4]{y#3}[#1][#2]}
\newcommandx{\zRel}[4][1=, 2=, 3=, 4=]{\mthrel[#4]{z#3}[#1][#2]}

\newcommandx{\ASym}[4][1=, 2=, 3=, 4=]{\mthsym[#4]{A#3}[#1][#2]}
\newcommandx{\BSym}[4][1=, 2=, 3=, 4=]{\mthsym[#4]{B#3}[#1][#2]}
\newcommandx{\CSym}[4][1=, 2=, 3=, 4=]{\mthsym[#4]{C#3}[#1][#2]}
\newcommandx{\DSym}[4][1=, 2=, 3=, 4=]{\mthsym[#4]{D#3}[#1][#2]}
\newcommandx{\ESym}[4][1=, 2=, 3=, 4=]{\mthsym[#4]{E#3}[#1][#2]}
\newcommandx{\FSym}[4][1=, 2=, 3=, 4=]{\mthsym[#4]{F#3}[#1][#2]}
\newcommandx{\GSym}[4][1=, 2=, 3=, 4=]{\mthsym[#4]{G#3}[#1][#2]}
\newcommandx{\HSym}[4][1=, 2=, 3=, 4=]{\mthsym[#4]{H#3}[#1][#2]}
\newcommandx{\ISym}[4][1=, 2=, 3=, 4=]{\mthsym[#4]{I#3}[#1][#2]}
\newcommandx{\JSym}[4][1=, 2=, 3=, 4=]{\mthsym[#4]{J#3}[#1][#2]}
\newcommandx{\KSym}[4][1=, 2=, 3=, 4=]{\mthsym[#4]{K#3}[#1][#2]}
\newcommandx{\LSym}[4][1=, 2=, 3=, 4=]{\mthsym[#4]{L#3}[#1][#2]}
\newcommandx{\MSym}[4][1=, 2=, 3=, 4=]{\mthsym[#4]{M#3}[#1][#2]}
\newcommandx{\NSym}[4][1=, 2=, 3=, 4=]{\mthsym[#4]{N#3}[#1][#2]}
\newcommandx{\OSym}[4][1=, 2=, 3=, 4=]{\mthsym[#4]{O#3}[#1][#2]}
\newcommandx{\PSym}[4][1=, 2=, 3=, 4=]{\mthsym[#4]{P#3}[#1][#2]}
\newcommandx{\QSym}[4][1=, 2=, 3=, 4=]{\mthsym[#4]{Q#3}[#1][#2]}
\newcommandx{\RSym}[4][1=, 2=, 3=, 4=]{\mthsym[#4]{R#3}[#1][#2]}
\newcommandx{\SSym}[4][1=, 2=, 3=, 4=]{\mthsym[#4]{S#3}[#1][#2]}
\newcommandx{\TSym}[4][1=, 2=, 3=, 4=]{\mthsym[#4]{T#3}[#1][#2]}
\newcommandx{\USym}[4][1=, 2=, 3=, 4=]{\mthsym[#4]{U#3}[#1][#2]}
\newcommandx{\VSym}[4][1=, 2=, 3=, 4=]{\mthsym[#4]{V#3}[#1][#2]}
\newcommandx{\WSym}[4][1=, 2=, 3=, 4=]{\mthsym[#4]{W#3}[#1][#2]}
\newcommandx{\XSym}[4][1=, 2=, 3=, 4=]{\mthsym[#4]{X#3}[#1][#2]}
\newcommandx{\YSym}[4][1=, 2=, 3=, 4=]{\mthsym[#4]{Y#3}[#1][#2]}
\newcommandx{\ZSym}[4][1=, 2=, 3=, 4=]{\mthsym[#4]{Z#3}[#1][#2]}

\newcommandx{\aSym}[4][1=, 2=, 3=, 4=]{\mthsym[#4]{a#3}[#1][#2]}
\newcommandx{\bSym}[4][1=, 2=, 3=, 4=]{\mthsym[#4]{b#3}[#1][#2]}
\newcommandx{\cSym}[4][1=, 2=, 3=, 4=]{\mthsym[#4]{c#3}[#1][#2]}
\newcommandx{\dSym}[4][1=, 2=, 3=, 4=]{\mthsym[#4]{d#3}[#1][#2]}
\newcommandx{\eSym}[4][1=, 2=, 3=, 4=]{\mthsym[#4]{e#3}[#1][#2]}
\newcommandx{\fSym}[4][1=, 2=, 3=, 4=]{\mthsym[#4]{f#3}[#1][#2]}
\newcommandx{\gSym}[4][1=, 2=, 3=, 4=]{\mthsym[#4]{g#3}[#1][#2]}
\newcommandx{\hSym}[4][1=, 2=, 3=, 4=]{\mthsym[#4]{h#3}[#1][#2]}
\newcommandx{\iSym}[4][1=, 2=, 3=, 4=]{\mthsym[#4]{i#3}[#1][#2]}
\newcommandx{\jSym}[4][1=, 2=, 3=, 4=]{\mthsym[#4]{j#3}[#1][#2]}
\newcommandx{\kSym}[4][1=, 2=, 3=, 4=]{\mthsym[#4]{k#3}[#1][#2]}
\newcommandx{\lSym}[4][1=, 2=, 3=, 4=]{\mthsym[#4]{l#3}[#1][#2]}
\newcommandx{\mSym}[4][1=, 2=, 3=, 4=]{\mthsym[#4]{m#3}[#1][#2]}
\newcommandx{\nSym}[4][1=, 2=, 3=, 4=]{\mthsym[#4]{n#3}[#1][#2]}
\newcommandx{\oSym}[4][1=, 2=, 3=, 4=]{\mthsym[#4]{o#3}[#1][#2]}
\newcommandx{\pSym}[4][1=, 2=, 3=, 4=]{\mthsym[#4]{p#3}[#1][#2]}
\newcommandx{\qSym}[4][1=, 2=, 3=, 4=]{\mthsym[#4]{q#3}[#1][#2]}
\newcommandx{\rSym}[4][1=, 2=, 3=, 4=]{\mthsym[#4]{r#3}[#1][#2]}
\newcommandx{\sSym}[4][1=, 2=, 3=, 4=]{\mthsym[#4]{s#3}[#1][#2]}
\newcommandx{\tSym}[4][1=, 2=, 3=, 4=]{\mthsym[#4]{t#3}[#1][#2]}
\newcommandx{\uSym}[4][1=, 2=, 3=, 4=]{\mthsym[#4]{u#3}[#1][#2]}
\newcommandx{\vSym}[4][1=, 2=, 3=, 4=]{\mthsym[#4]{v#3}[#1][#2]}
\newcommandx{\wSym}[4][1=, 2=, 3=, 4=]{\mthsym[#4]{w#3}[#1][#2]}
\newcommandx{\xSym}[4][1=, 2=, 3=, 4=]{\mthsym[#4]{x#3}[#1][#2]}
\newcommandx{\ySym}[4][1=, 2=, 3=, 4=]{\mthsym[#4]{y#3}[#1][#2]}
\newcommandx{\zSym}[4][1=, 2=, 3=, 4=]{\mthsym[#4]{z#3}[#1][#2]}

\newcommandx{\AElm}[4][1=, 2=, 3=, 4=]{\mthelm[#4]{A#3}[#1][#2]}
\newcommandx{\BElm}[4][1=, 2=, 3=, 4=]{\mthelm[#4]{B#3}[#1][#2]}
\newcommandx{\CElm}[4][1=, 2=, 3=, 4=]{\mthelm[#4]{C#3}[#1][#2]}
\newcommandx{\DElm}[4][1=, 2=, 3=, 4=]{\mthelm[#4]{D#3}[#1][#2]}
\newcommandx{\EElm}[4][1=, 2=, 3=, 4=]{\mthelm[#4]{E#3}[#1][#2]}
\newcommandx{\FElm}[4][1=, 2=, 3=, 4=]{\mthelm[#4]{F#3}[#1][#2]}
\newcommandx{\GElm}[4][1=, 2=, 3=, 4=]{\mthelm[#4]{G#3}[#1][#2]}
\newcommandx{\HElm}[4][1=, 2=, 3=, 4=]{\mthelm[#4]{H#3}[#1][#2]}
\newcommandx{\IElm}[4][1=, 2=, 3=, 4=]{\mthelm[#4]{I#3}[#1][#2]}
\newcommandx{\JElm}[4][1=, 2=, 3=, 4=]{\mthelm[#4]{J#3}[#1][#2]}
\newcommandx{\KElm}[4][1=, 2=, 3=, 4=]{\mthelm[#4]{K#3}[#1][#2]}
\newcommandx{\LElm}[4][1=, 2=, 3=, 4=]{\mthelm[#4]{L#3}[#1][#2]}
\newcommandx{\MElm}[4][1=, 2=, 3=, 4=]{\mthelm[#4]{M#3}[#1][#2]}
\newcommandx{\NElm}[4][1=, 2=, 3=, 4=]{\mthelm[#4]{N#3}[#1][#2]}
\newcommandx{\OElm}[4][1=, 2=, 3=, 4=]{\mthelm[#4]{O#3}[#1][#2]}
\newcommandx{\PElm}[4][1=, 2=, 3=, 4=]{\mthelm[#4]{P#3}[#1][#2]}
\newcommandx{\QElm}[4][1=, 2=, 3=, 4=]{\mthelm[#4]{Q#3}[#1][#2]}
\newcommandx{\RElm}[4][1=, 2=, 3=, 4=]{\mthelm[#4]{R#3}[#1][#2]}
\newcommandx{\SElm}[4][1=, 2=, 3=, 4=]{\mthelm[#4]{S#3}[#1][#2]}
\newcommandx{\TElm}[4][1=, 2=, 3=, 4=]{\mthelm[#4]{T#3}[#1][#2]}
\newcommandx{\UElm}[4][1=, 2=, 3=, 4=]{\mthelm[#4]{U#3}[#1][#2]}
\newcommandx{\VElm}[4][1=, 2=, 3=, 4=]{\mthelm[#4]{V#3}[#1][#2]}
\newcommandx{\WElm}[4][1=, 2=, 3=, 4=]{\mthelm[#4]{W#3}[#1][#2]}
\newcommandx{\XElm}[4][1=, 2=, 3=, 4=]{\mthelm[#4]{X#3}[#1][#2]}
\newcommandx{\YElm}[4][1=, 2=, 3=, 4=]{\mthelm[#4]{Y#3}[#1][#2]}
\newcommandx{\ZElm}[4][1=, 2=, 3=, 4=]{\mthelm[#4]{Z#3}[#1][#2]}

\newcommandx{\aElm}[4][1=, 2=, 3=, 4=]{\mthelm[#4]{a#3}[#1][#2]}
\newcommandx{\bElm}[4][1=, 2=, 3=, 4=]{\mthelm[#4]{b#3}[#1][#2]}
\newcommandx{\cElm}[4][1=, 2=, 3=, 4=]{\mthelm[#4]{c#3}[#1][#2]}
\newcommandx{\dElm}[4][1=, 2=, 3=, 4=]{\mthelm[#4]{d#3}[#1][#2]}
\newcommandx{\eElm}[4][1=, 2=, 3=, 4=]{\mthelm[#4]{e#3}[#1][#2]}
\newcommandx{\fElm}[4][1=, 2=, 3=, 4=]{\mthelm[#4]{f#3}[#1][#2]}
\newcommandx{\gElm}[4][1=, 2=, 3=, 4=]{\mthelm[#4]{g#3}[#1][#2]}
\newcommandx{\hElm}[4][1=, 2=, 3=, 4=]{\mthelm[#4]{h#3}[#1][#2]}
\newcommandx{\iElm}[4][1=, 2=, 3=, 4=]{\mthelm[#4]{i#3}[#1][#2]}
\newcommandx{\jElm}[4][1=, 2=, 3=, 4=]{\mthelm[#4]{j#3}[#1][#2]}
\newcommandx{\kElm}[4][1=, 2=, 3=, 4=]{\mthelm[#4]{k#3}[#1][#2]}
\newcommandx{\lElm}[4][1=, 2=, 3=, 4=]{\mthelm[#4]{l#3}[#1][#2]}
\newcommandx{\mElm}[4][1=, 2=, 3=, 4=]{\mthelm[#4]{m#3}[#1][#2]}
\newcommandx{\nElm}[4][1=, 2=, 3=, 4=]{\mthelm[#4]{n#3}[#1][#2]}
\newcommandx{\oElm}[4][1=, 2=, 3=, 4=]{\mthelm[#4]{o#3}[#1][#2]}
\newcommandx{\pElm}[4][1=, 2=, 3=, 4=]{\mthelm[#4]{p#3}[#1][#2]}
\newcommandx{\qElm}[4][1=, 2=, 3=, 4=]{\mthelm[#4]{q#3}[#1][#2]}
\newcommandx{\rElm}[4][1=, 2=, 3=, 4=]{\mthelm[#4]{r#3}[#1][#2]}
\newcommandx{\sElm}[4][1=, 2=, 3=, 4=]{\mthelm[#4]{s#3}[#1][#2]}
\newcommandx{\tElm}[4][1=, 2=, 3=, 4=]{\mthelm[#4]{t#3}[#1][#2]}
\newcommandx{\uElm}[4][1=, 2=, 3=, 4=]{\mthelm[#4]{u#3}[#1][#2]}
\newcommandx{\vElm}[4][1=, 2=, 3=, 4=]{\mthelm[#4]{v#3}[#1][#2]}
\newcommandx{\wElm}[4][1=, 2=, 3=, 4=]{\mthelm[#4]{w#3}[#1][#2]}
\newcommandx{\xElm}[4][1=, 2=, 3=, 4=]{\mthelm[#4]{x#3}[#1][#2]}
\newcommandx{\yElm}[4][1=, 2=, 3=, 4=]{\mthelm[#4]{y#3}[#1][#2]}
\newcommandx{\zElm}[4][1=, 2=, 3=, 4=]{\mthelm[#4]{z#3}[#1][#2]}

\newcommand{\ie}
	{\txtabr{i.e.}\xspace}

\newcommandx{\defeq}
	{\ensuremath{\triangleq}}
\newcommandx{\seteq}
	{\ensuremath{:=}}

\newcommand{\tuple}[1]
	{\ensuremath{\!\argint{\langle}{#1}{\rangle}}}

\newcommand{\tupleb}[2]
	{\tuple{\argb{#1}{#2}}}
\newcommand{\tuplec}[3]
	{\tuple{\argc{#1}{#2}{#3}}}
\newcommand{\tupled}[4]
	{\tuple{\argd{#1}{#2}{#3}{#4}}}
\newcommand{\tuplee}[5]
	{\tuple{\arge{#1}{#2}{#3}{#4}{#5}}}
\newcommand{\tuplef}[6]
	{\tuple{\argf{#1}{#2}{#3}{#4}{#5}{#6}}}
\newcommand{\tupleg}[7]
	{\tuple{\argg{#1}{#2}{#3}{#4}{#5}{#6}{#7}}}
\newcommand{\tupleh}[8]
	{\tuple{\argh{#1}{#2}{#3}{#4}{#5}{#6}{#7}{#8}}}
\newcommand{\tuplei}[9]
	{\tuple{\argi{#1}{#2}{#3}{#4}{#5}{#6}{#7}{#8}{#9}}}

\newcommand{\tuplecx}[3]
	{%
	\def\defarga{#1}%
	\def\defargb{#2}%
	\def\defargc{#3}%
	\argsubsup{\tupleauxcx}%
	}
\newcommand{\tupledx}[4]
	{%
	\def\defarga{#1}%
	\def\defargb{#2}%
	\def\defargc{#3}%
	\def\defargd{#4}%
	\argsubsup{\tupleauxdx}%
	}
\newcommand{\tupleex}[5]
	{%
	\def\defarga{#1}%
	\def\defargb{#2}%
	\def\defargc{#3}%
	\def\defargd{#4}%
	\def\defarge{#5}%
	\argsubsup{\tupleauxex}%
	}
\newcommand{\tuplefx}[6]
	{%
	\def\defarga{#1}%
	\def\defargb{#2}%
	\def\defargc{#3}%
	\def\defargd{#4}%
	\def\defarge{#5}%
	\def\defargf{#6}%
	\argsubsup{\tupleauxfx}%
	}
\newcommand{\tuplegx}[7]
	{%
	\def\defarga{#1}%
	\def\defargb{#2}%
	\def\defargc{#3}%
	\def\defargd{#4}%
	\def\defarge{#5}%
	\def\defargf{#6}%
	\def\defargg{#7}%
	\argsubsup{\tupleauxgx}%
	}

\newcommandx{\tupleauxbx}[2][1=, 2=]
	{%
	\tupleb
		{\argdef{#1}{\defarga[\argsubscript][\argsuperscript]}}
		{\argdef{#2}{\defargb[\argsubscript][\argsuperscript]}}%
	}
\newcommandx{\tupleauxcx}[3][1=, 2=, 3=]
	{%
	\tuplec
		{\argdef{#1}{\defarga[\argsubscript][\argsuperscript]}}
		{\argdef{#2}{\defargb[\argsubscript][\argsuperscript]}}
		{\argdef{#3}{\defargc[\argsubscript][\argsuperscript]}}%
	}
\newcommandx{\tupleauxdx}[4][1=, 2=, 3=, 4=]
	{%
	\tupled
		{\argdef{#1}{\defarga[\argsubscript][\argsuperscript]}}
		{\argdef{#2}{\defargb[\argsubscript][\argsuperscript]}}
		{\argdef{#3}{\defargc[\argsubscript][\argsuperscript]}}
		{\argdef{#4}{\defargd[\argsubscript][\argsuperscript]}}%
	}
\newcommandx{\tupleauxex}[5][1=, 2=, 3=, 4=, 5=]
	{%
	\tuplee
		{\argdef{#1}{\defarga[\argsubscript][\argsuperscript]}}
		{\argdef{#2}{\defargb[\argsubscript][\argsuperscript]}}
		{\argdef{#3}{\defargc[\argsubscript][\argsuperscript]}}
		{\argdef{#4}{\defargd[\argsubscript][\argsuperscript]}}
		{\argdef{#5}{\defarge[\argsubscript][\argsuperscript]}}%
	}
\newcommandx{\tupleauxfx}[6][1=, 2=, 3=, 4=, 5=, 6=]
	{%
	\tuplef
		{\argdef{#1}{\defarga[\argsubscript][\argsuperscript]}}
		{\argdef{#2}{\defargb[\argsubscript][\argsuperscript]}}
		{\argdef{#3}{\defargc[\argsubscript][\argsuperscript]}}
		{\argdef{#4}{\defargd[\argsubscript][\argsuperscript]}}
		{\argdef{#5}{\defarge[\argsubscript][\argsuperscript]}}
		{\argdef{#6}{\defargf[\argsubscript][\argsuperscript]}}%
	}
\newcommandx{\tupleauxgx}[7][1=, 2=, 3=, 4=, 5=, 6=, 7=]
	{%
	\tupleg
		{\argdef{#1}{\defarga[\argsubscript][\argsuperscript]}}
		{\argdef{#2}{\defargb[\argsubscript][\argsuperscript]}}
		{\argdef{#3}{\defargc[\argsubscript][\argsuperscript]}}
		{\argdef{#4}{\defargd[\argsubscript][\argsuperscript]}}
		{\argdef{#5}{\defarge[\argsubscript][\argsuperscript]}}
		{\argdef{#6}{\defargf[\argsubscript][\argsuperscript]}}
		{\argdef{#7}{\defargg[\argsubscript][\argsuperscript]}}%
	}
\newcommandx{\tupleauxhx}[8][1=, 2=, 3=, 4=, 5=, 6=, 7=, 8=]
	{%
	\tupleh
		{\argdef{#1}{\defarga[\argsubscript][\argsuperscript]}}
		{\argdef{#2}{\defargb[\argsubscript][\argsuperscript]}}
		{\argdef{#3}{\defargc[\argsubscript][\argsuperscript]}}
		{\argdef{#4}{\defargd[\argsubscript][\argsuperscript]}}
		{\argdef{#5}{\defarge[\argsubscript][\argsuperscript]}}
		{\argdef{#6}{\defargf[\argsubscript][\argsuperscript]}}
		{\argdef{#7}{\defargg[\argsubscript][\argsuperscript]}}
		{\argdef{#8}{\defargh[\argsubscript][\argsuperscript]}}%
	}
\newcommandx{\tupleauxix}[9][1=, 2=, 3=, 4=, 5=, 6=, 7=, 8=, 9=]
	{%
	\tuplei
		{\argdef{#1}{\defarga[\argsubscript][\argsuperscript]}}
		{\argdef{#2}{\defargb[\argsubscript][\argsuperscript]}}
		{\argdef{#3}{\defargc[\argsubscript][\argsuperscript]}}
		{\argdef{#4}{\defargd[\argsubscript][\argsuperscript]}}
		{\argdef{#5}{\defarge[\argsubscript][\argsuperscript]}}
		{\argdef{#6}{\defargf[\argsubscript][\argsuperscript]}}
		{\argdef{#7}{\defargg[\argsubscript][\argsuperscript]}}
		{\argdef{#8}{\defargh[\argsubscript][\argsuperscript]}}
		{\argdef{#9}{\defargi[\argsubscript][\argsuperscript]}}%
	}

\newcommand{\set}[2]
	{\ensuremath{\argint{\{}{\argext{#1}{\allowbreak:\allowbreak}{#2}}{\}}}}

\newcommand{\dom}
	{\mthargfun{dom}}

\newcommandx{\pto}[2][1=, 2=]
	{\ensuremath{\rightharpoonup}}

\newcommandx{\cto}[2][1=, 2=]
	{\:\mthempty{\to}[#1][#2]\:}
\newcommandx{\cpto}[2][1=, 2=]
	{\:\mthempty{\pto}[#1][#2]\:}

\newcommand{\numcc}[2]
	{\mthempty{[\argb{#1}{#2}]}}

\newcommand{\floor}[1]
	{\mthempty{\left\lfloor{#1}\right\rfloor}}

\newcommand{\argset}{Ar}
\newcommandx{\ArgSet}[3][1=, 2=, 3=]
	{\mthset{\argset#3}[#1][#2]}

\newcommand{\argsym}{a}
\newcommandx{\argSym}[3][1=, 2=, 3=]
	{\mthsym{\argsym#3}[#1][#2]}

\newcommand{\argelm}{a}
\newcommandx{\argElm}[3][1=, 2=, 3=]
	{\mthelm{\argelm#3}[#1][#2]}

\newcommand{\relset}{Rl}
\newcommandx{\RelSet}[3][1=, 2=, 3=]
	{\mthset{\relset#3}[#1][#2]}

\newcommand{\relsym}{r}
\newcommandx{\relSym}[3][1=, 2=, 3=]
	{\mthsym{\relsym#3}[#1][#2]}

\newcommand{\relelm}{r}
\newcommandx{\relElm}[3][1=, 2=, 3=]
	{\mthelm{\relelm#3}[#1][#2]}

\newcommand{\argfun}{ar}
\newcommandx{\argFun}[4][1=, 2=, 3=, 4=]
	{\mthargfun{\argfun#4}[#1][#2]{#3}}

\newcommand{\lansig}{LS}
\newcommandx{\LanSig}[5][1=, 2=, 3=, 4=, 5=]
	{\txtargname{\lansig#5{\small\argint{$[$}{#1}{$]$}}}[#2][#3]{#4}\xspace}

\newcommand{\lansigcls}{LS}
\newcommandx{\LanSigCls}[5][1=, 2=, 3=, 4=, 5=]
	{\mthset[#5]{\lansigcls#4\text{\txtname{\small\argint{$[$}{#1}{$]$}}}}[#2]%
	[#3]}

\newcommand{\domset}{Dm}
\newcommandx{\DomSet}[3][1=, 2=, 3=]
	{\mthset{\domset#3}[#1][#2]}

\newcommand{\domsym}{d}
\newcommandx{\domSym}[3][1=, 2=, 3=]
	{\mthsym{\domsym#3}[#1][#2]}

\newcommand{\domelm}{d}
\newcommandx{\domElm}[3][1=, 2=, 3=]
	{\mthelm{\domelm#3}[#1][#2]}

\newcommand{\relfun}{rl}
\newcommandx{\relFun}[4][1=, 2=, 3=, 4=]
	{\mthargfun{\relfun#4}[#1][#2]{#3}}

\newcommand{\relstr}{RS}
\newcommandx{\RelStr}[5][1=, 2=, 3=, 4=, 5=]
	{\txtargname{\relstr#5{\small\argint{$[$}{#1}{$]$}}}[#2][#3]{#4}\xspace}

\newcommand{\relstrcls}{RS}
\newcommandx{\RelStrCls}[5][1=, 2=, 3=, 4=, 5=]
	{\mthset[#5]{\relstrcls#4\text{\txtname{\small\argint{$[$}{#1}{$]$}}}}[#2]%
	[#3]}

\newcommandx{\ordFun}[3][1=, 2=, 3=]
	{\mthempty{\argint{\left\vert}{#3}{\right\vert}}[#1][#2]}

\newcommandx{\sizFun}[3][1=, 2=, 3=]
	{\mthempty{\argint{\left\Vert}{#3}{\right\Vert}}[#1][#2]}

\newcommand{\verset}{Vr}
\newcommandx{\VerSet}[3][1=, 2=, 3=]
	{\mthset{\verset#3}[#1][#2]}

\newcommand{\versym}{v}
\newcommandx{\verSym}[3][1=, 2=, 3=]
	{\mthsym{\versym#3}[#1][#2]}

\newcommand{\verelm}{v}
\newcommandx{\verElm}[3][1=, 2=, 3=]
	{\mthelm{\verelm#3}[#1][#2]}

\newcommand{\edgrel}{Ed}
\newcommandx{\EdgRel}[3][1=, 2=, 3=]
	{\mthrel{\edgrel#3}[#1][#2]}

\newcommand{\edgsym}{e}
\newcommandx{\edgSym}[3][1=, 2=, 3=]
	{\mthsym{\edgsym#3}[#1][#2]}

\newcommand{\edgelm}{e}
\newcommandx{\edgElm}[3][1=, 2=, 3=]
	{\mthelm{\edgelm#3}[#1][#2]}

\newcommand{\orgfun}{or}
\newcommandx{\orgFun}[4][1=, 2=, 3=, 4=]
	{\mthargfun{\orgfun#4}[#1][#2]{#3}}

\newcommand{\desfun}{ds}
\newcommandx{\desFun}[4][1=, 2=, 3=, 4=]
	{\mthargfun{\desfun#4}[#1][#2]{#3}}

\newcommand{\grp}{Gr}
\newcommandx{\Grp}[5][1=, 2=, 3=, 4=, 5=]
	{\txtargname{\grp#5{\small\argint{$[$}{#1}{$]$}}}[#2][#3]{#4}\xspace}

\newcommand{\grpcls}{Gr}
\newcommandx{\GrpCls}[5][1=, 2=, 3=, 4=, 5=]
	{\mthset[#5]{\grpcls#4\text{\small\txtname{\argint{$[$}{#1}{$]$}}}}[#2][#3]}

\newcommand{\pthset}{Pth}
\newcommandx{\PthSet}[3][1=, 2=, 3=]
	{\mthset{\pthset#3}[#1][#2]}

\newcommand{\pthsym}{\pi}
\newcommandx{\pthSym}[3][1=, 2=, 3=]
	{\mthsym{\pthsym#3}[#1][#2]}

\newcommand{\pthelm}{\pi}
\newcommandx{\pthElm}[3][1=, 2=, 3=]
	{\mthelm{\pthelm#3}[#1][#2]}

\newcommand{\colset}{Cl}
\newcommandx{\ColSet}[3][1=, 2=, 3=]
	{\mthset{\colset#3}[#1][#2]}

\newcommand{\colsym}{c}
\newcommandx{\colSym}[3][1=, 2=, 3=]
	{\mthsym{\colsym#3}[#1][#2]}

\newcommand{\colelm}{c}
\newcommandx{\colElm}[3][1=, 2=, 3=]
	{\mthelm{\colelm#3}[#1][#2]}

\newcommand{\colfun}{cl}
\newcommandx{\colFun}[4][1=, 2=, 3=, 4=]
	{\mthargfun{\colfun#4}[#1][#2]{#3}}

\newcommandx{\ColGrpCls}[5][1=, 2=, 3=, 4=, 5=]
	{\mthset[#5]{C}\!\GrpCls[#1][#2][#3][#4][#5]}

\newcommand{\wghset}{Wg}
\newcommandx{\WghSet}[3][1=, 2=, 3=]
	{\mthset{\wghset#3}[#1][#2]}

\newcommand{\wghsym}{w}
\newcommandx{\wghSym}[3][1=, 2=, 3=]
	{\mthsym{\wghsym#3}[#1][#2]}

\newcommand{\wghelm}{w}
\newcommandx{\wghElm}[3][1=, 2=, 3=]
	{\mthelm{\wghelm#3}[#1][#2]}

\newcommand{\wghfun}{wg}
\newcommandx{\wghFun}[4][1=, 2=, 3=, 4=]
	{\mthargfun{\wghfun#4}[#1][#2]{#3}}

\newcommandx{\WghGrpCls}[5][1=, 2=, 3=, 4=, 5=]
	{\mthset[#5]{W}\!\GrpCls[#1][#2][#3][#4][#5]}

\newcommand{\gamkin}{2PT}

\newcommand{\plrset}{Pl}
\newcommandx{\PlrSet}[3][1=, 2=, 3=]
	{\mthset{\plrset#3}[#1][#2]}

\newcommand{\plrsym}{p}
\newcommandx{\plrSym}[3][1=, 2=, 3=]
	{\mthsym{\plrsym#3}[#1][#2]}

\newcommand{\plrelm}{p}
\newcommandx{\plrElm}[3][1=, 2=, 3=]
	{\mthelm{\plrelm#3}[#1][#2]}

\newcommand{\agnset}{Ag}
\newcommandx{\AgnSet}[3][1=, 2=, 3=]
	{\mthset{\agnset#3}[#1][#2]}

\newcommand{\agnsym}{a}
\newcommandx{\agnSym}[3][1=, 2=, 3=]
	{\mthsym{\agnsym#3}[#1][#2]}

\newcommand{\agnelm}{a}
\newcommandx{\agnElm}[3][1=, 2=, 3=]
	{\mthelm{\agnelm#3}[#1][#2]}

\newcommand{\movset}{Mv}
\newcommandx{\MovSet}[3][1=, 2=, 3=]
	{\mthset{\movset#3}[#1][#2]}

\newcommand{\movrel}{Mv}
\newcommandx{\MovRel}[3][1=, 2=, 3=]
	{\mthrel{\movrel#3}[#1][#2]}

\newcommand{\movsym}{m}
\newcommandx{\movSym}[3][1=, 2=, 3=]
	{\mthsym{\movsym#3}[#1][#2]}

\newcommand{\movelm}{m}
\newcommandx{\movElm}[3][1=, 2=, 3=]
	{\mthelm{\movelm#3}[#1][#2]}

\newcommand{\actset}{Ac}
\newcommandx{\ActSet}[3][1=, 2=, 3=]
	{\mthset{\actset#3}[#1][#2]}

\newcommand{\actrel}{Ac}
\newcommandx{\ActRel}[3][1=, 2=, 3=]
	{\mthrel{\actrel#3}[#1][#2]}

\newcommand{\actsym}{c}
\newcommandx{\actSym}[3][1=, 2=, 3=]
	{\mthsym{\actsym#3}[#1][#2]}

\newcommand{\actelm}{c}
\newcommandx{\actElm}[3][1=, 2=, 3=]
	{\mthelm{\actelm#3}[#1][#2]}

\newcommand{\decset}{Dc}
\newcommandx{\DecSet}[3][1=, 2=, 3=]
	{\mthset{\decset#3}[#1][#2]}

\newcommand{\decsym}{\delta}
\newcommandx{\decSym}[4][1=, 2=, 3=, 4=]
	{\mthargfun{\decsym#4}[#1][#2]{#3}}

\newcommand{\decelm}{\delta}
\newcommandx{\decElm}[4][1=, 2=, 3=, 4=]
	{\mthargfun{\decelm#4}[#1][#2]{#3}}

\newcommand{\posset}{Ps}
\newcommandx{\PosSet}[3][1=, 2=, 3=]
	{\mthset{\posset#3}[#1][#2]}

\newcommand{\possym}{v}
\newcommandx{\posSym}[3][1=, 2=, 3=]
	{\mthsym{\possym#3}[#1][#2]}

\newcommand{\poselm}{v}
\newcommandx{\posElm}[3][1=, 2=, 3=]
	{\mthelm{\poselm#3}[#1][#2]}

\newcommand{\sttset}{St}
\newcommandx{\SttSet}[3][1=, 2=, 3=]
	{\mthset{\sttset#3}[#1][#2]}

\newcommand{\sttsym}{s}
\newcommandx{\sttSym}[3][1=, 2=, 3=]
	{\mthsym{\sttsym#3}[#1][#2]}

\newcommand{\sttelm}{s}
\newcommandx{\sttElm}[3][1=, 2=, 3=]
	{\mthelm{\sttelm#3}[#1][#2]}

\newcommand{\plrfun}{pl}
\newcommandx{\plrFun}[4][1=, 2=, 3=, 4=]
	{\mthargfun{\plrfun#4}[#1][#2]{#3}}

\newcommand{\agnfun}{ag}
\newcommandx{\agnFun}[4][1=, 2=, 3=, 4=]
	{\mthargfun{\agnfun#4}[#1][#2]{#3}}

\newcommand{\movfun}{mv}
\newcommandx{\movFun}[4][1=, 2=, 3=, 4=]
	{\mthargfun{\movfun#4}[#1][#2]{#3}}

\newcommand{\actfun}{ac}
\newcommandx{\actFun}[4][1=, 2=, 3=, 4=]
	{\mthargfun{\actfun#4}[#1][#2]{#3}}

\newcommand{\decfun}{dc}
\newcommandx{\decFun}[4][1=, 2=, 3=, 4=]
	{\mthargfun{\decfun#4}[#1][#2]{#3}}

\newcommand{\trnfun}{tr}
\newcommandx{\trnFun}[4][1=, 2=, 3=, 4=]
	{\mthargfun{\trnfun#4}[#1][#2]{#3}}

\newcommand{\arn}{Ar}
\newcommandx{\Arn}[5][1=, 2=, 3=, 4=, 5=]
	{\txtargname{\arn#5{\small\argint{$[$}{#1}{$]$}}}[#2][#3]{#4}\xspace}

\newcommand{\arnname}{A}
\newcommand{\ArnName}
	{\mthname{\arnname}}

\newcommand{\arncls}{Ar}
\newcommandx{\ArnCls}[5][1=, 2=, 3=, 4=, 5=]
	{\mthset[#5]{\arncls#4\text{\small\txtname{\argint{$[$}{#1}{$]$}}}}[#2][#3]}

\newcommand{\ArnStr}[1][]
	{%
	\IfStrEqCase{\argdef{#1}{\gamkin}}
		{%
		{2PT}
			{\tuplecx{\PosSet[0]}{\PosSet[1]}{\MovRel}}%
		{MPC0}
			{\tupledx{\PlrSet}{\MovSet}{\PosSet}{\trnFun}}%
		{MPC1}
			{\tupleex{\PlrSet}{\MovSet}{\PosSet}{\decFun}{\trnFun}}%
		{MPC2}
			{\tuplefx{\PlrSet}{\MovSet}{\PosSet}{\plrFun}{\movFun}{\trnFun}}%
		{MPC3}
			{\tuplegx{\PlrSet}{\MovSet}{\PosSet}{\plrFun}{\movFun}{\decFun}{\trnFun}}%
		{2AT}
			{\tuplecx{\SttSet[0]}{\SttSet[1]}{\ActRel}}%
		{MAC0}
			{\tupledx{\AgnSet}{\ActSet}{\SttSet}{\trnFun}}%
		{MAC1}
			{\tupleex{\AgnSet}{\ActSet}{\SttSet}{\decFun}{\trnFun}}%
		{MAC2}
			{\tuplefx{\AgnSet}{\ActSet}{\SttSet}{\agnFun}{\actFun}{\trnFun}}%
		{MAC3}
			{\tuplegx{\AgnSet}{\ActSet}{\SttSet}{\agnFun}{\actFun}{\decFun}{\trnFun}}%
		}
		[\ensuremath{\clubsuit}]%
	}

\newcommand{\hstset}{Hst}
\newcommandx{\HstSet}[3][1=, 2=, 3=]
	{\mthset{\hstset#3}[#1][#2]}

\newcommand{\hstsym}{\rho}
\newcommandx{\hstSym}[3][1=, 2=, 3=]
	{\mthsym{\hstsym#3}[#1][#2]}

\newcommand{\hstelm}{\rho}
\newcommandx{\hstElm}[3][1=, 2=, 3=]
	{\mthelm{\hstelm#3}[#1][#2]}

\newcommand{\strset}{Str}
\newcommandx{\StrSet}[3][1=, 2=, 3=]
	{\mthset{\strset#3}[#1][#2]}

\newcommand{\strsym}{\sigma}
\newcommandx{\strSym}[4][1=, 2=, 3=, 4=]
	{\mthargfun{\strsym#4}[#1][#2]{#3}}

\newcommand{\strelm}{\sigma}
\newcommandx{\strElm}[4][1=, 2=, 3=, 4=]
	{\mthargfun{\strelm#4}[#1][#2]{#3}}

\newcommand{\prfset}{Prf}
\newcommandx{\PrfSet}[3][1=, 2=, 3=]
	{\mthset{\prfset#3}[#1][#2]}

\newcommand{\prfsym}{\xi}
\newcommandx{\prfSym}[4][1=, 2=, 3=, 4=]
	{\mthargfun{\prfsym#4}[#1][#2]{#3}}

\newcommandx{\prfElm}[4][1=, 2=, 3=, 4=]
	{\mthargfun{\prfsym#4}[#1][#2]{#3}}

\newcommand{\playfun}{play}
\newcommandx{\playFun}[4][1=, 2=, 3=, 4=]
	{\mthargfun{\playfun#4}[#1][#2]{#3}}

\newcommandx{\ColArnCls}[5][1=, 2=, 3=, 4=, 5=]
	{\mthset[#5]{C}\!\ArnCls[#1][#2][#3][#4][#5]}

\newcommandx{\WghArnCls}[5][1=, 2=, 3=, 4=, 5=]
	{\mthset[#5]{W}\!\ArnCls[#1][#2][#3][#4][#5]}

\newcommand{\winset}{Wn}
\newcommandx{\WinSet}[3][1=, 2=, 3=]
	{\mthset{\winset#3}[#1][#2]}

\newcommand{\prdset}{Pr}
\newcommandx{\PrdSet}[3][1=, 2=, 3=]
	{\mthset{\prdset#3}[#1][#2]}

\newcommand{\prdsym}{p}
\newcommandx{\prdSym}[3][1=, 2=, 3=]
	{\mthsym{\prdsym#3}[#1][#2]}

\newcommand{\prdelm}{p}
\newcommandx{\prdElm}[3][1=, 2=, 3=]
	{\mthelm{\prdelm#3}[#1][#2]}

\newcommand{\prdfun}{pr}
\newcommandx{\prdFun}[4][1=, 2=, 3=, 4=]
	{\mthargfun{\prdfun#4}[#1][#2]{#3}}

\newcommand{\extname}{E}
\newcommand{\ExtName}
	{\mthname{\extname}}

\newcommand{\extcls}{Ex}
\newcommandx{\ExtCls}[5][1=, 2=, 3=, 4=, 5=]
	{\mthset[#5]{\extcls#4\text{\small\txtname{\argint{$[$}{#1}{$]$}}}}[#2][#3]}

\newcommand{\tarset}{Tr}
\newcommandx{\TarSet}[3][1=, 2=, 3=]
	{\mthset{\tarset#3}[#1][#2]}

\newcommand{\tarsym}{t}
\newcommandx{\tarSym}[3][1=, 2=, 3=]
	{\mthsym{\tarsym#3}[#1][#2]}

\newcommand{\tarelm}{t}
\newcommandx{\tarElm}[3][1=, 2=, 3=]
	{\mthelm{\tarelm#3}[#1][#2]}

\newcommand{\schrel}{\models}
\newcommandx{\schRel}[4][1=, 2=, 3=, 4=]
	{\mthrel{\schrel#3}[#1][#2]}

\newcommand{\schcls}{Sc}
\newcommandx{\SchCls}[5][1=, 2=, 3=, 4=, 5=]
	{\mthset[#5]{\schcls#4\text{\small\txtname{\argint{$[$}{#1}{$]$}}}}[#2][#3]}

\newcommand{\gamcls}{Gm}
\newcommandx{\GamCls}[5][1=, 2=, 3=, 4=, 5=]
	{\mthset[#5]{\gamcls#4\text{\small\txtname{\argint{$[$}{#1}{$]$}}}}[#2][#3]}

\newcommandx{\GamStr}[3][1=, 2=, 3=]
	{%
	\StrLeft{\argdef{#1}{\gamkin}}{2}[\optgamkin]%
	\def\defposelm{#2}%
	\def\deftarelm{#3}%
	\argsubsup{\GamStrAux}%
	}
\newcommandx{\GamStrAux}[2][1=, 2=]
	{%
	\IfStrEqCase{\optgamkin}
		{%
		{2P}
			{%
			\tuplec
				{\argdef{#1}{\ArnName[\argsubscript][\argsuperscript]}}
				{\argdef{\defposelm}{\posElm[\argsubscript][\argsuperscript]}}
				{\argdef{#2}{\WinSet[\argsubscript][\argsuperscript]}}%
			}%
		{MP}
			{%
			\tuplec
				{\argdef{#1}{\ExtName[\argsubscript][\argsuperscript]}}
				{\argdef{\defposelm}{\posElm[\argsubscript][\argsuperscript]}}
				{\argdef{\deftarelm}{\tarElm[\argsubscript][\argsuperscript]}}%
			}%
		{2A}
			{%
			\tuplec
				{\argdef{#1}{\ArnName[\argsubscript][\argsuperscript]}}
				{\argdef{\defposelm}{\sttElm[\argsubscript][\argsuperscript]}}
				{\argdef{#2}{\WinSet[\argsubscript][\argsuperscript]}}%
			}%
		{MA}
			{%
			\tuplec
				{\argdef{#1}{\ExtName[\argsubscript][\argsuperscript]}}
				{\argdef{\defposelm}{\sttElm[\argsubscript][\argsuperscript]}}
				{\argdef{\deftarelm}{\tarElm[\argsubscript][\argsuperscript]}}%
			}%
		}
		[\ensuremath{\clubsuit}]%
	}

\newcommand{\apset}{AP}
\newcommandx{\APSet}[3][1=, 2=, 3=]
	{\mthset{\apset#3}[#1][#2]}

\newcommand{\apsym}{p}
\newcommandx{\apSym}[3][1=, 2=, 3=]
	{\mthsym{\apsym#3}[#1][#2]}

\newcommand{\apelm}{p}
\newcommandx{\apElm}[3][1=, 2=, 3=]
	{\mthelm{\apelm#3}[#1][#2]}

\newcommand{\worset}{W}
\newcommandx{\WorSet}[3][1=, 2=, 3=]
	{\mthset{\worset#3}[#1][#2]}

\newcommand{\worsym}{w}
\newcommandx{\worSym}[3][1=, 2=, 3=]
	{\mthsym{\worsym#3}[#1][#2]}

\newcommand{\worelm}{w}
\newcommandx{\worElm}[3][1=, 2=, 3=]
	{\mthelm{\worelm#3}[#1][#2]}

\newcommand{\trnrel}{R}
\newcommandx{\TrnRel}[3][1=, 2=, 3=]
	{\mthrel{\trnrel#3}[#1][#2]}

\newcommand{\trnsym}{r}
\newcommandx{\trnSym}[3][1=, 2=, 3=]
	{\mthsym{\trnsym#3}[#1][#2]}

\newcommand{\trnelm}{r}
\newcommandx{\trnElm}[3][1=, 2=, 3=]
	{\mthelm{\trnelm#3}[#1][#2]}

\newcommand{\labfun}{L}
\newcommandx{\labFun}[4][1=, 2=, 3=, 4=]
	{\mthargfun{\labfun#4}[#1][#2]{#3}}

\newcommand{\krpstr}{KS}
\newcommandx{\KrpStr}[5][1=, 2=, 3=, 4=, 5=]
	{\txtargname{\krpstr#5{\small\argint{$[$}{#1}{$]$}}}[#2][#3]{#4}\xspace}

\newcommand{\krpstrcls}{KS}
\newcommandx{\KrpStrCls}[5][1=, 2=, 3=, 4=, 5=]
	{\mthset[#5]{\krpstrcls#4\text{\small\txtname{\argint{$[$}{#1}{$]$}}}}[#2]%
	[#3]}

\newcommand{\trkset}{Trk}
\newcommandx{\TrkSet}[3][1=, 2=, 3=]
	{\mthset{\trkset#3}[#1][#2]}

\newcommand{\trksym}{\rho}
\newcommandx{\trkSym}[3][1=, 2=, 3=]
	{\mthsym{\trksym#3}[#1][#2]}

\newcommand{\trkelm}{\rho}
\newcommandx{\trkElm}[3][1=, 2=, 3=]
	{\mthelm{\trkelm#3}[#1][#2]}

\newcommand{\krptree}{KT}
\newcommandx{\KrpTree}[5][1=, 2=, 3=, 4=, 5=]
	{\txtargname{\krptree#5{\small\argint{$[$}{#1}{$]$}}}[#2][#3]{#4}\xspace}

\newcommand{\krptreecls}{KT}
\newcommandx{\KrpTreeCls}[5][1=, 2=, 3=, 4=, 5=]
	{\mthset[#5]{\krptreecls#4\text{\small\txtname{\argint{$[$}{#1}{$]$}}}}[#2]%
	[#3]}

\newcommand{\dirset}{Dir}
\newcommandx{\DirSet}[3][1=, 2=, 3=]
	{\mthset{\dirset#3}[#1][#2]}

\newcommand{\dirsym}{d}
\newcommandx{\dirSym}[3][1=, 2=, 3=]
	{\mthsym{\dirsym#3}[#1][#2]}

\newcommand{\direlm}{d}
\newcommandx{\dirElm}[3][1=, 2=, 3=]
	{\mthelm{\direlm#3}[#1][#2]}

\newcommand{\unwfun}{unw}
\newcommandx{\unwFun}[4][1=, 2=, 3=, 4=]
	{\mthargfun{\unwfun#4}[#1][#2]{#3}}

\newcommand{\congamstr}{CGS}
\newcommandx{\ConGamStr}[5][1=, 2=, 3=, 4=, 5=]
	{\txtargname{\congamstr#5{\small\argint{$[$}{#1}{$]$}}}[#2][#3]{#4}\xspace}

\newcommand{\trntabkin}{D}

\newcommand{\symset}{Sm}
\newcommandx{\SymSet}[3][1=, 2=, 3=]
	{\mthset{\symset#3}[#1][#2]}

\newcommand{\symsym}{\ell}
\newcommandx{\symSym}[3][1=, 2=, 3=]
	{\mthsym{\symsym#3}[#1][#2]}

\newcommand{\symelm}{\ell}
\newcommandx{\symElm}[3][1=, 2=, 3=]
	{\mthelm{\symelm#3}[#1][#2]}

\newcommand{\DSttSet}[1][]
	{\SttSet[\Delta#1]}

\newcommand{\ESttSet}[1][]
	{\SttSet[\exists#1]}

\newcommand{\ASttSet}[1][]
	{\SttSet[\forall#1]}

\newcommand{\trntab}{tt}
\newcommandx{\TrnTab}[5][1=, 2=, 3=, 4=, 5=]
	{\txtargname{\trntab#5{\small\argint{$[$}{#1}{$]$}}}[#2][#3]{#4}\xspace}

\newcommand{\trntabcls}{TT}
\newcommandx{\TrnTabCls}[5][1=, 2=, 3=, 4=, 5=]
	{\mthset[#5]{\trntabcls#4\text{\txtname{\small\argint{$[$}{#1}{$]$}}}}[#2]%
	[#3]}

\newcommand{\TrnTabStr}[1][]
	{%
	\IfStrEqCase{\argdef{#1}{\trntabkin}}
		{%
		{D}{\tuplecx{\SymSet}{\SttSet}{\trnFun}}%
		{N}{\tupledx{\SymSet}{\DSttSet}{\ESttSet}{\trnFun}}%
		{U}{\tupledx{\SymSet}{\DSttSet}{\ASttSet}{\trnFun}}%
		{A}{\tupleex{\SymSet}{\DSttSet}{\ESttSet}{\ASttSet}{\trnFun}}%
		}
		[\ensuremath{\clubsuit}]%
	}

\newcommandx{\FOL}[5][1=, 2=, 3=, 4=, 5=]
	{\txtargname{FOL#5{\small\argint{$[$}{#1}{$]$}}}[#2][#3]{#4}\xspace}

\newcommand{\varset}{Vr}
\newcommandx{\VarSet}[3][1=, 2=, 3=]
	{\mthset{\varset#3}[#1][#2]}

\newcommand{\varsym}{x}
\newcommandx{\varSym}[3][1=, 2=, 3=]
	{\mthsym{\varsym#3}[#1][#2]}

\newcommand{\varelm}{x}
\newcommandx{\varElm}[3][1=, 2=, 3=]
	{\mthelm{\varelm#3}[#1][#2]}

\newcommand{\varfun}{vr}
\newcommandx{\varFun}[4][1=, 2=, 3=, 4=]
	{\mthargfun{\varfun#4}[#1][#2]{#3}}

\newcommand{\freeFun}
	{\mthargfun{free}}

\newcommand{\qntset}{Qn}
\newcommandx{\QntSet}[3][1=, 2=, 3=]
	{\mthset{\qntset#3}[#1][#2]}

\newcommand{\qntsym}{\wp}
\newcommandx{\qntSym}[3][1=, 2=, 3=]
	{\mthsym{\qntsym#3}[#1][#2]}

\newcommand{\qntelm}{\wp}
\newcommandx{\qntElm}[3][1=, 2=, 3=]
	{\mthelm{\qntelm#3}[#1][#2]}

\newcommand{\bndset}{Bn}
\newcommandx{\BndSet}[3][1=, 2=, 3=]
	{\mthset{\bndset#3}[#1][#2]}

\newcommand{\bndsym}{\flat}
\newcommandx{\bndSym}[3][1=, 2=, 3=]
	{\mthsym{\bndsym#3}[#1][#2]}

\newcommand{\bndelm}{\flat}
\newcommandx{\bndElm}[3][1=, 2=, 3=]
	{\mthelm{\bndelm#3}[#1][#2]}

\newcommand{\depset}{\Delta}
\newcommandx{\DepSet}[3][1=, 2=, 3=]
	{\mthset{\depset#3}[#1][#2]}

\newcommand{\asgset}{Asg}
\newcommandx{\AsgSet}[3][1=, 2=, 3=]
	{\mthset{\asgset#3}[#1][#2]}

\newcommand{\asgfun}{\chi}
\newcommandx{\asgFun}[4][1=, 2=, 3=, 4=]
	{\mthargfun{\asgfun#4}[#1][#2]{#3}}

\newcommand{\smset}{SM}
\newcommandx{\SMSet}[3][1=, 2=, 3=]
	{\mthset{\smset#3}[#1][#2]}

\newcommand{\smfun}{\delta}
\newcommandx{\smFun}[4][1=, 2=, 3=, 4=]
	{\mthargfun{\smfun#4}[#1][#2]{#3}}

\newcommand{\cmset}{CM}
\newcommandx{\CMSet}[3][1=, 2=, 3=]
	{\mthset{\cmset#3}[#1][#2]}

\newcommand{\cmfun}{\gamma}
\newcommandx{\cmFun}[4][1=, 2=, 3=, 4=]
	{\mthargfun{\cmfun#4}[#1][#2]{#3}}

\newcommand{\schset}{Sch}
\newcommandx{\SchSet}[3][1=, 2=, 3=]
	{\mthset{\schset#3}[#1][#2]}

\newcommand{\schsym}{\sigma}
\newcommandx{\schSym}[3][1=, 2=, 3=]
	{\mthsym{\schsym#3}[#1][#2]}

\newcommand{\schelm}{\sigma}
\newcommandx{\schElm}[3][1=, 2=, 3=]
	{\mthelm{\schelm#3}[#1][#2]}

\newcommand{\entset}{Ent}
\newcommandx{\EntSet}[4][1=, 2=, 3=, 4=]
	{\mthset{\entset#4}[#1][#2]{#3}}

\newcommand{\entfun}{ent}
\newcommandx{\entFun}[4][1=, 2=, 3=, 4=]
	{\mthargfun{\entfun#4}[#1][#2]{#3}}

\newcommandx{\SOL}[5][1=, 2=, 3=, 4=, 5=]
	{\txtargname{SOL#5{\small\argint{$[$}{#1}{$]$}}}[#2][#3]{#4}\xspace}

\newcommandx{\TL}[5][1=, 2=, 3=, 4=, 5=]
	{\txtargname{TL#5{\small\argint{$[$}{#1}{$]$}}}[#2][#3]{#4}\xspace}

\newcommandx{\PL}[5][1=, 2=, 3=, 4=, 5=]
	{\txtargname{PL#5{\small\argint{$[$}{#1}{$]$}}}[#2][#3]{#4}\xspace}

\newcommand{\fvarset}{FVr}
\newcommandx{\FVarSet}[3][1=, 2=, 3=]
	{\mthset{\fvarset#3}[#1][#2]}

\newcommand{\fvarsym}{x}
\newcommandx{\fvarSym}[3][1=, 2=, 3=]
	{\mthsym{\fvarsym#3}[#1][#2]}

\newcommand{\fvarelm}{x}
\newcommandx{\fvarElm}[3][1=, 2=, 3=]
	{\mthelm{\fvarelm#3}[#1][#2]}

\newcommand{\fvarfun}{fvr}
\newcommandx{\fvarFun}[4][1=, 2=, 3=, 4=]
	{\mthargfun{\fvarfun#4}[#1][#2]{#3}}

\newcommand{\svarset}{SVr}
\newcommandx{\SVarSet}[3][1=, 2=, 3=]
	{\mthset{\svarset#3}[#1][#2]}

\newcommand{\svarsym}{X}
\newcommandx{\svarSym}[3][1=, 2=, 3=]
	{\mthsym{\svarsym#3}[#1][#2]}

\newcommand{\svarelm}{X}
\newcommandx{\svarElm}[3][1=, 2=, 3=]
	{\mthelm{\svarelm#3}[#1][#2]}

\newcommand{\svarfun}{svr}
\newcommandx{\svarFun}[4][1=, 2=, 3=, 4=]
	{\mthargfun{\svarfun#4}[#1][#2]{#3}}

\newcommandx{\MuCalculus}[5][1=, 2=, 3=, 4=, 5=]
	{\txtargname{$\mu$Calculus#5{\small\argint{$[$}{#1}{$]$}}}[#2][#3]{#4}\xspace}

\newcommandx{\LTL}[5][1=, 2=, 3=, 4=, 5=]
	{\txtargname{LTL#5{\small\argint{$[$}{#1}{$]$}}}[#2][#3]{#4}\xspace}

\newcommandx{\PTL}[5][1=, 2=, 3=, 4=, 5=]
	{\txtargname{PTL#5{\small\argint{$[$}{#1}{$]$}}}[#2][#3]{#4}\xspace}

\newcommand{\X}
	{\mthsym{X}}

\newcommand{\F}
	{\mthsym{F}}

\newcommand{\G}
	{\mthsym{G}}

\newcommand{\U}
	{\mthsym{U}}

\newcommand{\B}
	{\mthsym{B}}

\newcommandx{\CTL}[5][1=, 2=, 3=, 4=, 5=]
	{\txtargname{CTL#5{\small\argint{$[$}{#1}{$]$}}}[#2][#3]{#4}\xspace}

\newcommandx{\CTLP}[5][1=, 2=, 3=, 4=, 5=]
	{\txtargname{CTL$^{+}$#5{\small\argint{$[$}{#1}{$]$}}}[#2][#3]{#4}\xspace}

\newcommandx{\CTLS}[5][1=, 2=, 3=, 4=, 5=]
	{\txtargname{CTL$^{\star}$#5{\small\argint{$[$}{#1}{$]$}}}[#2][#3]{#4}\xspace}

\newcommand{\A}
	{\mthsym{A}}

\newcommandx{\STL}[5][1=, 2=, 3=, 4=, 5=]
	{\txtargname{STL#5{\small\argint{$[$}{#1}{$]$}}}[#2][#3]{#4}\xspace}

\newcommandx{\STLP}[5][1=, 2=, 3=, 4=, 5=]
	{\txtargname{STL$^{+}$#5{\small\argint{$[$}{#1}{$]$}}}[#2][#3]{#4}\xspace}

\newcommandx{\STLS}[5][1=, 2=, 3=, 4=, 5=]
	{\txtargname{STL$^{\star}$#5{\small\argint{$[$}{#1}{$]$}}}[#2][#3]{#4}\xspace}

\newcommandx{\ATL}[5][1=, 2=, 3=, 4=, 5=]
	{\txtargname{ATL#5{\small\argint{$[$}{#1}{$]$}}}[#2][#3]{#4}\xspace}

\newcommandx{\ATLP}[5][1=, 2=, 3=, 4=, 5=]
	{\txtargname{ATL$^{+}$#5{\small\argint{$[$}{#1}{$]$}}}[#2][#3]{#4}\xspace}

\newcommandx{\ATLS}[5][1=, 2=, 3=, 4=, 5=]
	{\txtargname{ATL$^{\star}$#5{\small\argint{$[$}{#1}{$]$}}}[#2][#3]{#4}\xspace}

\newcommandx{\SL}[5][1=, 2=, 3=, 4=, 5=]
	{\txtargname{SL#5{\small\argint{$[$}{#1}{$]$}}}[#2][#3]{#4}\xspace}

\newcommand{\EExs}[1]
	{\ensuremath{%
	\argint{\mbox{$\langle\!\langle$}}{#1}{\mbox{$\rangle\!\rangle$}}%
	}}

\newcommand{\AAll}[1]
	{\ensuremath{\argint{\mbox{$[\:\!\![$}}{#1}{\mbox{$]\:\!\!]$}}}}

\newcommandx{\EF}[5][1=, 2=, 3=, 4=, 5=]
	{\txtargname{EF#5{\small\argint{$[$}{#1}{$]$}}}[#2][#3]{#4}\xspace}

\newcommandx{\SG}[5][1=, 2=, 3=, 4=, 5=]
	{\txtargname{SG#5{\small\argint{$[$}{#1}{$]$}}}[#2][#3]{#4}\xspace}

\newcommandx{\LogTime}[4][1=, 2=, 3=, 4=]
	{\txtargname{LogTime#4}[#2][#3]{#1}\xspace}
\newcommandx{\LogTimeH}[4][1=, 2=, 3=, 4=]
	{\LogTime[#1][#2][#3][#4]-\HComp}
\newcommandx{\LogTimeE}[4][1=, 2=, 3=, 4=]
	{\LogTime[#1][#2][#3][#4]-\EComp}
\newcommandx{\LogTimeC}[4][1=, 2=, 3=, 4=]
	{\LogTime[#1][#2][#3][#4]-\CComp}

\newcommand{\NLogTime}
	{\txtname{N}\LogTime}
\newcommandx{\NLogTimeH}[4][1=, 2=, 3=, 4=]
	{\NLogTime[#1][#2][#3][#4]-\HComp}
\newcommandx{\NLogTimeE}[4][1=, 2=, 3=, 4=]
	{\NLogTime[#1][#2][#3][#4]-\EComp}
\newcommandx{\NLogTimeC}[4][1=, 2=, 3=, 4=]
	{\NLogTime[#1][#2][#3][#4]-\CComp}

\newcommand{\CoNLogTime}
	{\txtname{Co}\NLogTime}
\newcommandx{\CoNLogTimeH}[4][1=, 2=, 3=, 4=]
	{\CoNLogTime[#1][#2][#3][#4]-\HComp}
\newcommandx{\CoNLogTimeE}[4][1=, 2=, 3=, 4=]
	{\CoNLogTime[#1][#2][#3][#4]-\EComp}
\newcommandx{\CoNLogTimeC}[4][1=, 2=, 3=, 4=]
	{\CoNLogTime[#1][#2][#3][#4]-\CComp}

\newcommandx{\ALogTime}
	{\txtname{A}\LogTime}
\newcommandx{\ALogTimeH}[4][1=, 2=, 3=, 4=]
	{\ALogTime[#1][#2][#3][#4]-\HComp}
\newcommandx{\ALogTimeE}[4][1=, 2=, 3=, 4=]
	{\ALogTime[#1][#2][#3][#4]-\EComp}
\newcommandx{\ALogTimeC}[4][1=, 2=, 3=, 4=]
	{\ALogTime[#1][#2][#3][#4]-\CComp}

\newcommandx{\LogSpace}[4][1=, 2=, 3=, 4=]
	{\txtargname{LogSpace#4}[#2][#3]{#1}\xspace}
\newcommandx{\LogSpaceH}[4][1=, 2=, 3=, 4=]
	{\LogSpace[#1][#2][#3][#4]-\HComp}
\newcommandx{\LogSpaceE}[4][1=, 2=, 3=, 4=]
	{\LogSpace[#1][#2][#3][#4]-\EComp}
\newcommandx{\LogSpaceC}[4][1=, 2=, 3=, 4=]
	{\LogSpace[#1][#2][#3][#4]-\CComp}

\newcommandx{\NLogSpace}
	{\txtname{N}\LogSpace}
\newcommandx{\NLogSpaceH}[4][1=, 2=, 3=, 4=]
	{\NLogSpace[#1][#2][#3][#4]-\HComp}
\newcommandx{\NLogSpaceE}[4][1=, 2=, 3=, 4=]
	{\NLogSpace[#1][#2][#3][#4]-\EComp}
\newcommandx{\NLogSpaceC}[4][1=, 2=, 3=, 4=]
	{\NLogSpace[#1][#2][#3][#4]-\CComp}

\newcommandx{\CoNLogSpace}
	{\txtname{Co}\NLogSpace}
\newcommandx{\CoNLogSpaceH}[4][1=, 2=, 3=, 4=]
	{\CoNLogSpace[#1][#2][#3][#4]-\HComp}
\newcommandx{\CoNLogSpaceE}[4][1=, 2=, 3=, 4=]
	{\CoNLogSpace[#1][#2][#3][#4]-\EComp}
\newcommandx{\CoNLogSpaceC}[4][1=, 2=, 3=, 4=]
	{\CoNLogSpace[#1][#2][#3][#4]-\CComp}

\newcommandx{\ALogSpace}
	{\txtname{A}\LogSpace}
\newcommandx{\ALogSpaceH}[4][1=, 2=, 3=, 4=]
	{\ALogSpace[#1][#2][#3][#4]-\HComp}
\newcommandx{\ALogSpaceE}[4][1=, 2=, 3=, 4=]
	{\ALogSpace[#1][#2][#3][#4]-\EComp}
\newcommandx{\ALogSpaceC}[4][1=, 2=, 3=, 4=]
	{\ALogSpace[#1][#2][#3][#4]-\CComp}

\newcommandx{\PTime}[4][1=, 2=, 3=, 4=]
	{\txtargname{PTime#4}[#2][#3]{#1}\xspace}
\newcommandx{\PTimeH}[4][1=, 2=, 3=, 4=]
	{\PTime[#1][#2][#3][#4]-\HComp}
\newcommandx{\PTimeE}[4][1=, 2=, 3=, 4=]
	{\PTime[#1][#2][#3][#4]-\EComp}
\newcommandx{\PTimeC}[4][1=, 2=, 3=, 4=]
	{\PTime[#1][#2][#3][#4]-\CComp}

\newcommandx{\UPTime}
	{\txtname{U}\PTime}
\newcommandx{\UPTimeH}[4][1=, 2=, 3=, 4=]
	{\UPTime[#1][#2][#3][#4]-\HComp}
\newcommandx{\UPTimeE}[4][1=, 2=, 3=, 4=]
	{\UPTime[#1][#2][#3][#4]-\EComp}
\newcommandx{\UPTimeC}[4][1=, 2=, 3=, 4=]
	{\UPTime[#1][#2][#3][#4]-\CComp}

\newcommandx{\CoUPTime}
	{\txtname{Co}\UPTime}
\newcommandx{\CoUPTimeH}[4][1=, 2=, 3=, 4=]
	{\CoUPTime[#1][#2][#3][#4]-\HComp}
\newcommandx{\CoUPTimeE}[4][1=, 2=, 3=, 4=]
	{\CoUPTime[#1][#2][#3][#4]-\EComp}
\newcommandx{\CoUPTimeC}[4][1=, 2=, 3=, 4=]
	{\CoUPTime[#1][#2][#3][#4]-\CComp}

\newcommandx{\NPTime}
	{\txtname{N}\PTime}
\newcommandx{\NPTimeH}[4][1=, 2=, 3=, 4=]
	{\NPTime[#1][#2][#3][#4]-\HComp}
\newcommandx{\NPTimeE}[4][1=, 2=, 3=, 4=]
	{\NPTime[#1][#2][#3][#4]-\EComp}
\newcommandx{\NPTimeC}[4][1=, 2=, 3=, 4=]
	{\NPTime[#1][#2][#3][#4]-\CComp}

\newcommandx{\CoNPTime}
	{\txtname{Co}\NPTime}
\newcommandx{\CoNPTimeH}[4][1=, 2=, 3=, 4=]
	{\CoNPTime[#1][#2][#3][#4]-\HComp}
\newcommandx{\CoNPTimeE}[4][1=, 2=, 3=, 4=]
	{\CoNPTime[#1][#2][#3][#4]-\EComp}
\newcommandx{\CoNPTimeC}[4][1=, 2=, 3=, 4=]
	{\CoNPTime[#1][#2][#3][#4]-\CComp}

\newcommandx{\APTime}
	{\txtname{A}\PTime}
\newcommandx{\APTimeH}[4][1=, 2=, 3=, 4=]
	{\APTime[#1][#2][#3][#4]-\HComp}
\newcommandx{\APTimeE}[4][1=, 2=, 3=, 4=]
	{\APTime[#1][#2][#3][#4]-\EComp}
\newcommandx{\APTimeC}[4][1=, 2=, 3=, 4=]
	{\APTime[#1][#2][#3][#4]-\CComp}

\newcommandx{\PSpace}[4][1=, 2=, 3=, 4=]
	{\txtargname{PSpace#4}[#2][#3]{#1}\xspace}
\newcommandx{\PSpaceH}[4][1=, 2=, 3=, 4=]
	{\PSpace[#1][#2][#3][#4]-\HComp}
\newcommandx{\PSpaceE}[4][1=, 2=, 3=, 4=]
	{\PSpace[#1][#2][#3][#4]-\EComp}
\newcommandx{\PSpaceC}[4][1=, 2=, 3=, 4=]
	{\PSpace[#1][#2][#3][#4]-\CComp}

\newcommandx{\NPSpace}
	{\txtname{N}\PSpace}
\newcommandx{\NPSpaceH}[4][1=, 2=, 3=, 4=]
	{\NPSpace[#1][#2][#3][#4]-\HComp}
\newcommandx{\NPSpaceE}[4][1=, 2=, 3=, 4=]
	{\NPSpace[#1][#2][#3][#4]-\EComp}
\newcommandx{\NPSpaceC}[4][1=, 2=, 3=, 4=]
	{\NPSpace[#1][#2][#3][#4]-\CComp}

\newcommandx{\CoNPSpace}
	{\txtname{Co}\NPSpace}
\newcommandx{\CoNPSpaceH}[4][1=, 2=, 3=, 4=]
	{\CoNPSpace[#1][#2][#3][#4]-\HComp}
\newcommandx{\CoNPSpaceE}[4][1=, 2=, 3=, 4=]
	{\CoNPSpace[#1][#2][#3][#4]-\EComp}
\newcommandx{\CoNPSpaceC}[4][1=, 2=, 3=, 4=]
	{\CoNPSpace[#1][#2][#3][#4]-\CComp}

\newcommandx{\APSpace}
	{\txtname{A}\PSpace}
\newcommandx{\APSpaceH}[4][1=, 2=, 3=, 4=]
	{\APSpace[#1][#2][#3][#4]-\HComp}
\newcommandx{\APSpaceE}[4][1=, 2=, 3=, 4=]
	{\APSpace[#1][#2][#3][#4]-\EComp}
\newcommandx{\APSpaceC}[4][1=, 2=, 3=, 4=]
	{\APSpace[#1][#2][#3][#4]-\CComp}

\newcommandx{\ExpTime}[4][1=, 2=, 3=, 4=]
	{\txtargname{ExpTime#4}[#2][#3]{#1}\xspace}
\newcommandx{\ExpTimeH}[4][1=, 2=, 3=, 4=]
	{\ExpTime[#1][#2][#3][#4]-\HComp}
\newcommandx{\ExpTimeE}[4][1=, 2=, 3=, 4=]
	{\ExpTime[#1][#2][#3][#4]-\EComp}
\newcommandx{\ExpTimeC}[4][1=, 2=, 3=, 4=]
	{\ExpTime[#1][#2][#3][#4]-\CComp}

\newcommandx{\NExpTime}
	{\txtname{N}\ExpTime}
\newcommandx{\NExpTimeH}[4][1=, 2=, 3=, 4=]
	{\NExpTime[#1][#2][#3][#4]-\HComp}
\newcommandx{\NExpTimeE}[4][1=, 2=, 3=, 4=]
	{\NExpTime[#1][#2][#3][#4]-\EComp}
\newcommandx{\NExpTimeC}[4][1=, 2=, 3=, 4=]
	{\NExpTime[#1][#2][#3][#4]-\CComp}

\newcommandx{\CoNExpTime}
	{\txtname{Co}\NExpTime}
\newcommandx{\CoNExpTimeH}[4][1=, 2=, 3=, 4=]
	{\CoNExpTime[#1][#2][#3][#4]-\HComp}
\newcommandx{\CoNExpTimeE}[4][1=, 2=, 3=, 4=]
	{\CoNExpTime[#1][#2][#3][#4]-\EComp}
\newcommandx{\CoNExpTimeC}[4][1=, 2=, 3=, 4=]
	{\CoNExpTime[#1][#2][#3][#4]-\CComp}

\newcommandx{\AExpTime}
	{\txtname{A}\ExpTime}
\newcommandx{\AExpTimeH}[4][1=, 2=, 3=, 4=]
	{\AExpTime[#1][#2][#3][#4]-\HComp}
\newcommandx{\AExpTimeE}[4][1=, 2=, 3=, 4=]
	{\AExpTime[#1][#2][#3][#4]-\EComp}
\newcommandx{\AExpTimeC}[4][1=, 2=, 3=, 4=]
	{\AExpTime[#1][#2][#3][#4]-\CComp}

\newcommandx{\ExpSpace}[4][1=, 2=, 3=, 4=]
	{\txtargname{ExpSpace#4}[#2][#3]{#1}\xspace}
\newcommandx{\ExpSpaceH}[4][1=, 2=, 3=, 4=]
	{\ExpSpace[#1][#2][#3][#4]-\HComp}
\newcommandx{\ExpSpaceE}[4][1=, 2=, 3=, 4=]
	{\ExpSpace[#1][#2][#3][#4]-\EComp}
\newcommandx{\ExpSpaceC}[4][1=, 2=, 3=, 4=]
	{\ExpSpace[#1][#2][#3][#4]-\CComp}

\newcommandx{\NExpSpace}
	{\txtname{N}\ExpSpace}
\newcommandx{\NExpSpaceH}[4][1=, 2=, 3=, 4=]
	{\NExpSpace[#1][#2][#3][#4]-\HComp}
\newcommandx{\NExpSpaceE}[4][1=, 2=, 3=, 4=]
	{\NExpSpace[#1][#2][#3][#4]-\EComp}
\newcommandx{\NExpSpaceC}[4][1=, 2=, 3=, 4=]
	{\NExpSpace[#1][#2][#3][#4]-\CComp}

\newcommandx{\CoNExpSpace}
	{\txtname{Co}\NExpSpace}
\newcommandx{\CoNExpSpaceH}[4][1=, 2=, 3=, 4=]
	{\CoNExpSpace[#1][#2][#3][#4]-\HComp}
\newcommandx{\CoNExpSpaceE}[4][1=, 2=, 3=, 4=]
	{\CoNExpSpace[#1][#2][#3][#4]-\EComp}
\newcommandx{\CoNExpSpaceC}[4][1=, 2=, 3=, 4=]
	{\CoNExpSpace[#1][#2][#3][#4]-\CComp}

\newcommandx{\AExpSpace}
	{\txtname{A}\ExpSpace}
\newcommandx{\AExpSpaceH}[4][1=, 2=, 3=, 4=]
	{\AExpSpace[#1][#2][#3][#4]-\HComp}
\newcommandx{\AExpSpaceE}[4][1=, 2=, 3=, 4=]
	{\AExpSpace[#1][#2][#3][#4]-\EComp}
\newcommandx{\AExpSpaceC}[4][1=, 2=, 3=, 4=]
	{\AExpSpace[#1][#2][#3][#4]-\CComp}

\newcommandx{\NonElmTime}[4][1=, 2=, 3=, 4=]
	{\txtargname{NonElementaryTime#4}[#2][#3]{#1}\xspace}
\newcommandx{\NonElmTimeH}[4][1=, 2=, 3=, 4=]
	{\NonElmTime[#1][#2][#3][#4]-\HComp}
\newcommandx{\NonElmTimeE}[4][1=, 2=, 3=, 4=]
	{\NonElmTime[#1][#2][#3][#4]-\EComp}
\newcommandx{\NonElmTimeC}[4][1=, 2=, 3=, 4=]
	{\NonElmTime[#1][#2][#3][#4]-\CComp}

\newcommandx{\NonElmSpace}[4][1=, 2=, 3=, 4=]
	{\txtargname{NonElementarySpace#4}[#2][#3]{#1}\xspace}
\newcommandx{\NonElmSpaceH}[4][1=, 2=, 3=, 4=]
	{\NonElmSpace[#1][#2][#3][#4]-\HComp}
\newcommandx{\NonElmSpaceE}[4][1=, 2=, 3=, 4=]
	{\NonElmSpace[#1][#2][#3][#4]-\EComp}
\newcommandx{\NonElmSpaceC}[4][1=, 2=, 3=, 4=]
	{\NonElmSpace[#1][#2][#3][#4]-\CComp}

\newcommandx{\DLHier}[4][2=, 3=, 4=]
	{\mthargset[0]{\Delta#4}[#1][#3]{#2}\xspace}
\newcommandx{\DLHierH}[4][2=, 3=, 4=]
	{\DLHier{#1}[#2][#3][#4]-\HComp}
\newcommandx{\DLHierE}[4][2=, 3=, 4=]
	{\DLHier{#1}[#2][#3][#4]-\EComp}
\newcommandx{\DLHierC}[4][2=, 3=, 4=]
	{\DLHier{#1}[#2][#3][#4]-\CComp}

\newcommandx{\ELHier}[4][2=, 3=, 4=]
	{\mthargset[0]{\Sigma#4}[#1][#3]{#2}\xspace}
\newcommandx{\ELHierH}[4][2=, 3=, 4=]
	{\ELHier{#1}[#2][#3][#4]-\HComp}
\newcommandx{\ELHierE}[4][2=, 3=, 4=]
	{\ELHier{#1}[#2][#3][#4]-\EComp}
\newcommandx{\ELHierC}[4][2=, 3=, 4=]
	{\ELHier{#1}[#2][#3][#4]-\CComp}

\newcommandx{\ULHier}[4][2=, 3=, 4=]
	{\mthargset[0]{\Pi#4}[#1][#3]{#2}\xspace}
\newcommandx{\ULHierH}[4][2=, 3=, 4=]
	{\ULHier{#1}[#2][#3][#4]-\HComp}
\newcommandx{\ULHierE}[4][2=, 3=, 4=]
	{\ULHier{#1}[#2][#3][#4]-\EComp}
\newcommandx{\ULHierC}[4][2=, 3=, 4=]
	{\ULHier{#1}[#2][#3][#4]-\CComp}

\newcommandx{\DBHier}[4][2=, 3=, 4=]
	{\mthargset[3]{\Delta#4}[#1][#3]{#2}\xspace}
\newcommandx{\DBHierH}[4][2=, 3=, 4=]
	{\DBHier{#1}[#2][#3][#4]-\HComp}
\newcommandx{\DBHierE}[4][2=, 3=, 4=]
	{\DBHier{#1}[#2][#3][#4]-\EComp}
\newcommandx{\DBHierC}[4][2=, 3=, 4=]
	{\DBHier{#1}[#2][#3][#4]-\CComp}

\newcommandx{\EBHier}[4][2=, 3=, 4=]
	{\mthargset[3]{\Sigma#4}[#1][#3]{#2}\xspace}
\newcommandx{\EBHierH}[4][2=, 3=, 4=]
	{\EBHier{#1}[#2][#3][#4]-\HComp}
\newcommandx{\EBHierE}[4][2=, 3=, 4=]
	{\EBHier{#1}[#2][#3][#4]-\EComp}
\newcommandx{\EBHierC}[4][2=, 3=, 4=]
	{\EBHier{#1}[#2][#3][#4]-\CComp}

\newcommandx{\UBHier}[4][2=, 3=, 4=]
	{\mthargset[3]{\Pi#4}[#1][#3]{#2}\xspace}
\newcommandx{\UBHierH}[4][2=, 3=, 4=]
	{\UBHier{#1}[#2][#3][#4]-\HComp}
\newcommandx{\UBHierE}[4][2=, 3=, 4=]
	{\UBHier{#1}[#2][#3][#4]-\EComp}
\newcommandx{\UBHierC}[4][2=, 3=, 4=]
	{\UBHier{#1}[#2][#3][#4]-\CComp}

\newcommandx{\DPolHier}[4][2=, 3=, 4=]
	{\DLHier{#1}[#2][\argb{\mathrm{P}}{#3}][#4]}
\newcommandx{\DPolHierH}[4][2=, 3=, 4=]
	{\DPolHier{#1}[#2][#3][#4]-\HComp}
\newcommandx{\DPolHierE}[4][2=, 3=, 4=]
	{\DPolHier{#1}[#2][#3][#4]-\EComp}
\newcommandx{\DPolHierC}[4][2=, 3=, 4=]
	{\DPolHier{#1}[#2][#3][#4]-\CComp}

\newcommandx{\EPolHier}[4][2=, 3=, 4=]
	{\ELHier{#1}[#2][\argb{\mathrm{P}}{#3}][#4]}
\newcommandx{\EPolHierH}[4][2=, 3=, 4=]
	{\EPolHier{#1}[#2][#3][#4]-\HComp}
\newcommandx{\EPolHierE}[4][2=, 3=, 4=]
	{\EPolHier{#1}[#2][#3][#4]-\EComp}
\newcommandx{\EPolHierC}[4][2=, 3=, 4=]
	{\EPolHier{#1}[#2][#3][#4]-\CComp}

\newcommandx{\UPolHier}[4][2=, 3=, 4=]
	{\ULHier{#1}[#2][\argb{\mathrm{P}}{#3}][#4]}
\newcommandx{\UPolHierH}[4][2=, 3=, 4=]
	{\UPolHier{#1}[#2][#3][#4]-\HComp}
\newcommandx{\UPolHierE}[4][2=, 3=, 4=]
	{\UPolHier{#1}[#2][#3][#4]-\EComp}
\newcommandx{\UPolHierC}[4][2=, 3=, 4=]
	{\UPolHier{#1}[#2][#3][#4]-\CComp}

\newcommandx{\DAriHier}[4][2=, 3=, 4=]
	{\DLHier{#1}[#2][\argb{0}{#3}][#4]}
\newcommandx{\DAriHierH}[4][2=, 3=, 4=]
	{\DAriHier{#1}[#2][#3][#4]-\HComp}
\newcommandx{\DAriHierE}[4][2=, 3=, 4=]
	{\DAriHier{#1}[#2][#3][#4]-\EComp}
\newcommandx{\DAriHierC}[4][2=, 3=, 4=]
	{\DAriHier{#1}[#2][#3][#4]-\CComp}

\newcommandx{\EAriHier}[4][2=, 3=, 4=]
	{\ELHier{#1}[#2][\argb{0}{#3}][#4]}
\newcommandx{\EAriHierH}[4][2=, 3=, 4=]
	{\EAriHier{#1}[#2][#3][#4]-\HComp}
\newcommandx{\EAriHierE}[4][2=, 3=, 4=]
	{\EAriHier{#1}[#2][#3][#4]-\EComp}
\newcommandx{\EAriHierC}[4][2=, 3=, 4=]
	{\EAriHier{#1}[#2][#3][#4]-\CComp}

\newcommandx{\UAriHier}[4][2=, 3=, 4=]
	{\ULHier{#1}[#2][\argb{0}{#3}][#4]}
\newcommandx{\UAriHierH}[4][2=, 3=, 4=]
	{\UAriHier{#1}[#2][#3][#4]-\HComp}
\newcommandx{\UAriHierE}[4][2=, 3=, 4=]
	{\UAriHier{#1}[#2][#3][#4]-\EComp}
\newcommandx{\UAriHierC}[4][2=, 3=, 4=]
	{\UAriHier{#1}[#2][#3][#4]-\CComp}

\newcommandx{\DAnaHier}[4][2=, 3=, 4=]
	{\DLHier{#1}[#2][\argb{1}{#3}][#4]}
\newcommandx{\DAnaHierH}[4][2=, 3=, 4=]
	{\DAnaHier{#1}[#2][#3][#4]-\HComp}
\newcommandx{\DAnaHierE}[4][2=, 3=, 4=]
	{\DAnaHier{#1}[#2][#3][#4]-\EComp}
\newcommandx{\DAnaHierC}[4][2=, 3=, 4=]
	{\DAnaHier{#1}[#2][#3][#4]-\CComp}

\newcommandx{\EAnaHier}[4][2=, 3=, 4=]
	{\ELHier{#1}[#2][\argb{1}{#3}][#4]}
\newcommandx{\EAnaHierH}[4][2=, 3=, 4=]
	{\EAnaHier{#1}[#2][#3][#4]-\HComp}
\newcommandx{\EAnaHierE}[4][2=, 3=, 4=]
	{\EAnaHier{#1}[#2][#3][#4]-\EComp}
\newcommandx{\EAnaHierC}[4][2=, 3=, 4=]
	{\EAnaHier{#1}[#2][#3][#4]-\CComp}

\newcommandx{\UAnaHier}[4][2=, 3=, 4=]
	{\ULHier{#1}[#2][\argb{1}{#3}][#4]}
\newcommandx{\UAnaHierH}[4][2=, 3=, 4=]
	{\UAnaHier{#1}[#2][#3][#4]-\HComp}
\newcommandx{\UAnaHierE}[4][2=, 3=, 4=]
	{\UAnaHier{#1}[#2][#3][#4]-\EComp}
\newcommandx{\UAnaHierC}[4][2=, 3=, 4=]
	{\UAnaHier{#1}[#2][#3][#4]-\CComp}

\newcommandx{\DBorHier}[4][2=, 3=, 4=]
	{\DBHier{#1}[#2][\argb{\mathrm{B}}{#3}][#4]}
\newcommandx{\DBorHierH}[4][2=, 3=, 4=]
	{\DBorHier{#1}[#2][#3][#4]-\HComp}
\newcommandx{\DBorHierE}[4][2=, 3=, 4=]
	{\DBorHier{#1}[#2][#3][#4]-\EComp}
\newcommandx{\DBorHierC}[4][2=, 3=, 4=]
	{\DBorHier{#1}[#2][#3][#4]-\CComp}

\newcommandx{\EBorHier}[4][2=, 3=, 4=]
	{\EBHier{#1}[#2][\argb{\mathrm{B}}{#3}][#4]}
\newcommandx{\EBorHierH}[4][2=, 3=, 4=]
	{\EBorHier{#1}[#2][#3][#4]-\HComp}
\newcommandx{\EBorHierE}[4][2=, 3=, 4=]
	{\EBorHier{#1}[#2][#3][#4]-\EComp}
\newcommandx{\EBorHierC}[4][2=, 3=, 4=]
	{\EBorHier{#1}[#2][#3][#4]-\CComp}

\newcommandx{\UBorHier}[4][2=, 3=, 4=]
	{\UBHier{#1}[#2][\argb{\mathrm{B}}{#3}][#4]}
\newcommandx{\UBorHierH}[4][2=, 3=, 4=]
	{\UBorHier{#1}[#2][#3][#4]-\HComp}
\newcommandx{\UBorHierE}[4][2=, 3=, 4=]
	{\UBorHier{#1}[#2][#3][#4]-\EComp}
\newcommandx{\UBorHierC}[4][2=, 3=, 4=]
	{\UBorHier{#1}[#2][#3][#4]-\CComp}

\newcommand{\HComp}
	{\txtname{hard}\xspace}

\newcommand{\EComp}
	{\txtname{easy}\xspace}

\newcommand{\CComp}
	{\txtname{complete}\xspace}

% ...

% ...

\newcommandx{\MCMAS}
	{MCMAS\xspace}

\newcommand{\MCMASSLK}
	{MCMAS-SLK\xspace}

\newcommand{\CTLK}
	{\CTL[][][][][K]}

\newcommand{\SLK}
	{\SL[][][][][K]}

\newcommand{\K}
	{\mthsym{K}}
\newcommand{\D}
	{\mthsym{D}}
\newcommand{\C}
	{\mthsym{C}}

\newcommand{\Tuple}[1]{\left\langle#1\right\rangle}

\renewcommand{\B}[2]{\left(#1,#2\right)\!}

\newcommand{\sharingFun}
	{\mthfun{sharing}}

\newcommandx{\ExtSet}[3][1=, 2=, 3=]
	{\mthset{Ext#3}[#1][#2]}

\renewcommandx{\BndSet}[3][1=, 2=, 3=]
	{\mthset{Bnd#3}[#1][#2]}

\newcommand{\SatFun}
	{\mthfun{Sat}}

% ...

\begin{document}

\title
	{\MCMASSLK: A Model Checker for the Verification of Strategy Logic
	Specifications}

\author
	{Petr \v{C}erm\'ak\inst{1} \and Alessio Lomuscio\inst{1} \and Fabio
	Mogavero\inst{2} \and Aniello Murano\inst{2}}

\institute
	{Imperial College London, UK \and Universit\`a degli Studi di Napoli Federico
	II, Italy}

\maketitle

%%****************************************************************************%%
%%                                                                            %%
%% A Model Checker for the Verification of Strategy Logic Specifications      %%
%%                                                                            %%
%% Introduction.tex                                                           %%
%%                                                                            %%
%% Revision 1                                                                 %%
%%                                                                            %%
%% Copyright (C) 2014, Petr Cermak, Alessio Lomuscio, Fabio Mogavero, and     %%
%%                     Aniello Murano.                                        %%
%% All rights reserved.                                                       %%
%%                                                                            %%
%%****************************************************************************%%

% Begin of file Introduction.tex

\begin{section}{Introduction}

Model checking has come of age.
A number of techniques are increasingly used in industrial setting to verify
hardware and software systems, both against models and concrete implementations.
While it is generally accepted that obstacles still remain, notably handling
infinite state systems efficiently, much of current work involves refining and
improving existing techniques such as predicate abstraction.

At scientific level a major avenue of work remains the development of
verification techniques against rich and expressive specification languages.
Over the years there has been a natural progression from checking reachability
only to a large number of techniques (BDDs, BMC, abstraction, etc.) catering
for \LTL~\cite{Pnu77}, \CTL~\cite{CE81}, and \CTLS~\cite{EH86}.
More recently, \ATL and \ATLS~\cite{AHK02} were introduced to analyse systems in
which some components, or \emph{agents}, can enforce temporal properties on the
system.
The paths so identified correspond to infinite games between a coalition and its
complement.
\ATL is well explored theoretically and at least two toolkits now support
it~\cite{AHMQRT98,LQR09,LR06a}.

It has however been observed that \ATLS suffers from a number of limitations
when one tries to apply it to multi-agent system reasoning and
games~\cite{AGJ07,WHW07,AW09,FS10, JM14, BLLM09,LLM10}.
One of these is the lack of support for binding strategies explicitly to various
agents or to the same agent in different contexts.
To overcome this and other difficulties, \emph{Strategy Logic}
(\SL)~\cite{MMV10b}, as well as some useful variants of
it~\cite{CHP10,MMPV12,MMS13a,MMS14}, has been put forward.
Key game-theoretic properties such as \emph{Nash equilibria}, not expressible in
\ATLS, can be captured in \SL.

In this paper we describe the model checker \MCMASSLK.
The tool supports the verification of systems against specifications expressed
in a variant of \SL that includes epistemic modalities.
The synthesis of agents' strategies to satisfy a given parametric specification,
as well as basic counterexample generation, are also supported.
\MCMASSLK, released as open-source, implements novel labelling algorithms for
\SL, encoded on BDDs, and reuses existing algorithms for the verification of
epistemic specifications~\cite{RL07}.

\end{section}

% End of file Introduction.tex

%%****************************************************************************%%
%%                                                                            %%
%% A Model Checker for the Verification of Strategy Logic Specifications      %%
%%                                                                            %%
%% SectionI.tex                                                               %%
%%                                                                            %%
%% Revision 1                                                                 %%
%%                                                                            %%
%% Copyright (C) 2014, Petr Cermak, Alessio Lomuscio, Fabio Mogavero, and     %%
%%                     Aniello Murano.                                        %%
%% All rights reserved.                                                       %%
%%                                                                            %%
%%****************************************************************************%%

% Begin of file SectionI.tex

\begin{section}{Epistemic Strategy Logic}
\label{sec:epsstrlog}

\textbf{Underlying Framework.}
Differently from other treatments of \SL, originally defined on concurrent game
structures, we here define the logic on \emph{interpreted
systems}~\cite{FHMV95}.
Doing so enables us to integrate the logic with epistemic concepts.
Each agent is modelled in terms of its local states (given as a set of
variables), a set of actions, a protocol specifying what actions may be
performed at a given local state, and a local evolution function returning a
target local state given a local state and a joint action for all the agents
in the system.
Interpreted systems are attractive for their modularity; they naturally express
systems with incomplete information, and are amenable to
verification~\cite{GM04,LQR09}.

\textbf{Syntax.}
\SL has been introduced as a powerful formalism to reason about various
equilibria concepts in non-zero sum games and sophisticated cooperation concepts
in multi-agent systems~\cite{CHP10,MMV10b}.
These are not expressible in previously explored logics including those in the
\ATLS hierarchy.
We here put forward an epistemic extension of \SL by adding a family of
knowledge operators~\cite{FHMV95}.

%% we consider here an extension of
%% \emph{Strategy Logic}~\cite{MMV10b} with knowledge operators (\SLK,
%% for short), a first-order logic over memoryless
%% strategies\footnote{Note that the memoryless semantics has been never
%%   consider along with \SL, which has been only studied \wrt memoryful
%%   strategies.}, where temporal goals of agents are described with
%% \LTL~\cite{Pnu77} formulas through suitable strategy quantifications
%% and agent bindings.

Formulas in epistemic strategy logic, or strategy logic with knowledge (\SLK),
are built by the following grammar over atomic propositions $\pElm \in \APSet$,
variables $\varElm \in \VarSet$, and agents $\aElm \in \AgnSet$ ($\ASet
\subseteq \AgnSet$ denotes a set of agents):
\begin{center}
	$\varphi ::= % \top \mid
	\pElm \mid \neg \varphi \mid \varphi \wedge \varphi \mid \X \varphi \mid
	\varphi \U \varphi \mid \EExs{\xElm} \varphi \mid (\aElm, \xElm) \varphi \mid
	\K[\aElm] \varphi \mid \D[\ASet] \varphi \mid \C[\ASet] \varphi$.
\end{center}
\SLK extends \LTL~\cite{Pnu77} by means of an \emph{existential
  strategy quantifier} $\EExs{\xElm}$ and \emph{agent binding}
$(\aElm, \xElm)$.  It also includes the epistemic operators
$\K[\aElm]$, $\D[\ASet]$, and $\C[\ASet]$ for individual, distributed,
and common knowledge, respectively~\cite{FHMV95}.  Intuitively,
$\EExs{\xElm} \varphi$ is read as \emph{``there exists a strategy
  $\xElm$ such that $\varphi$ holds''}, whereas $(\aElm, \xElm)
\varphi$ stands for \emph{``bind agent $\aElm$ to the strategy
  associated with the variable $\xElm$ in $\varphi$''}.  The epistemic
formula $\K[\aElm] \varphi$ stands for \emph{``agent $\aElm$ knows
  that $\varphi$''}; $\D[\ASet] \varphi$ encodes \emph{``the group
  $\ASet$ has distributed knowledge of $\varphi$''}; while $\C[\ASet]
\varphi$ represents \emph{``the group $\ASet$ has common knowledge of
  $\varphi$''}.
Similarly to first-order languages, we use $\freeFun{\varphi}$ to represent the
\emph{free agents and variables} in a formula $\varphi$.
Formally, $\freeFun{\varphi} \subseteq \AgnSet \cup \VarSet$ contains \emph{(i)}
all agents having no binding after the occurrence of a temporal operator and
\emph{(ii)} all variables having a binding but no quantification.
For simplicity, we here consider only formulas where the epistemic modalities
are applied to sentences, \ie, formulas without free agents or variables.
Lifting this restriction is not problematic.
To establish the truth of a formula, the set of strategies over which a variable
can range needs to be determined.
For this purpose we use the set $\sharingFun(\varphi, \xElm)$ containing all
agents bound to a variable $\xElm$ within a formula $\varphi$.

\textbf{Semantics.}
The concepts of \emph{path}, \emph{play}, \emph{strategy}, and \emph{assignment}
(for agents and variables) can be defined on interpreted systems similarly to
the way they are defined on concurrent game structures.
We refer to~\cite{MMV10b,MMPV11} for a detailed presentation.
Intuitively, a strategy identifies paths in the model on which a formula needs
to be verified.
Various variants of interpreted systems have been studied.
We here adopt the memoryless version where the agents' local states do not
necessarily include the local history of the run.
Consequently, strategies are also memoryless.
Note that this markedly differs from the previous perfect recall semantics of
\SL, which is defined on memoryful strategies.
We consider this setting because memoryful semantics with incomplete information
leads to an undecidable model checking problem~\cite{DT11}.

%% In order to have a computationally grounded semantics~\cite{Woo00} for
%% a wide family of MAS formalisms, \MCMAS uses a generalization of
%% \emph{interpreted systems}~\cite{FHMV95}, called ISPL~\cite{LR06a}, as
%% a semantic framework for its specification languages.  Here, following
%% the same approach and differently from the original definition of
%% \SL~\cite{MMV10b}, we describe the meaning of an \SLK formula \wrt
%% ISPLs too.  In particular, we exploit the distributed structure of
%% this model to define the semantics of the knowledge operators, in a
%% way that is similar to the approach used for \ATL in~\cite{LR06a}.
%% Indeed, each agent is characterized by a set of local states and
%% actions that are performed following an \apriori given local protocol.
%% The whole system evolves in compliance with an evolution function that
%% determines how the local state of an agent changes as a function of
%% its state and other agent actions.  On such a model, in a way similar
%% to what is done for concurrent game structures, it is possible to
%% define the concepts of \emph{path}, \emph{play}, \emph{strategy}, and
%% \emph{assignment} (for agents and variables), which are at the base of
%% the definition of \SLK semantics.  As usual, a strategy allows to
%% select a subset of paths from the model on which to verify temporal
%% properties of a formula.  Observe that, in our framework, strategies
%% are even simpler due to the memoryless feature we require.  We refer
%% to~\cite{MMV10b,MMPV11} for a detailed presentation of all these
%% concepts.

Given an interpreted system $\IName$ having $G$ as a set of global states, a
state $\gElm \in G$, and an assignment $\asgFun$ defined on $\freeFun{\varphi}$,
we write $\IName, \asgFun, \gElm \models \varphi$ to indicate that the \SLK
formula $\varphi$ holds at $\gElm$ in $\IName$ under $\asgFun$.
The semantics of \SLK formulas is inductively defined by using the usual \LTL
interpretation for the atomic propositions, the Boolean connectives $\neg$ and
$\wedge$, as well as the temporal operators $\X$ and $\U$.
The epistemic modalities are interpreted as standard by relying on notions of
equality on the underlying sets of local states~\cite{FHMV95}.
The inductive cases for strategy quantification $\EExs{\xElm}$ and agent binding
$(\aElm, \xElm)$ are given as follows.
$\IName, \asgFun, \gElm \models \EExs{\xElm} \varphi$ iff there is a memoryless
strategy $\fFun$ for the agents in $\sharingFun(\varphi, \xElm)$ such that
$\IName, \asgFun{[\xElm \mapsto \fFun]}, \gElm \models \varphi$ where
$\asgFun{[\xElm \mapsto \fFun]}$ is the assignment equal to $\asgFun$ except for
the variable $\xElm$, for which it assumes the value $\fFun$.
$\IName, \asgFun, \gElm \models (\xElm, \aElm) \varphi$ iff $\IName,
\asgFun{[\aElm \mapsto \asgFun(\xElm)]}, \gElm \models \varphi$, where
$\asgFun{[\aElm \mapsto \asgFun(\xElm)]}$ denotes the assignment $\asgFun$ in
which agent $\aElm$ is bound to the strategy $\asgFun(\xElm)$.

\textbf{Model Checking and Strategy Synthesis.}
Given an interpreted system $\IName$, an initial global state $\gElm_0$, an \SLK
specification $\varphi$, and an assignment $\asgFun$ defined on
$\freeFun{\varphi}$, the \emph{model checking problem} concerns determining
whether $\IName, \asgFun, \gElm_0 \models \varphi$.
Given an interpreted system $\IName$, an initial global state $\gElm_0$, and an
\SLK specification $\varphi$, the \emph{strategy synthesis problem} involves
finding an assignment $\asgFun$ such that $\IName, \asgFun, \gElm_0 \models
\varphi$.

The model checking problem for systems with memoryless strategies and imperfect
information against \ATL and \ATLS specifications is in \PSpace~\cite{BDJ10}.
The algorithm can be adapted to show that the same result applies to \SLK.
It follows that \SLK specifications do not generate a harder model checking
problem even though they are more expressive.
%%   has the same
%% complexity even though it is more expressive than \ATLS.  Indeed, \SLK
%% can describe Nash equilibria as we show in the following sections.
%% %% In addition, \wrt \ATL, our logic results to be more expressive, both from a
%% %% strategic and temporal point of view.

\end{section}

% End of file SectionI.tex

%%****************************************************************************%%
%%                                                                            %%
%% A Model Checker for the Verification of Strategy Logic Specifications      %%
%%                                                                            %%
%% SectionII.tex                                                              %%
%%                                                                            %%
%% Revision 1                                                                 %%
%%                                                                            %%
%% Copyright (C) 2014, Petr Cermak, Alessio Lomuscio, Fabio Mogavero, and     %%
%%                     Aniello Murano.                                        %%
%% All rights reserved.                                                       %%
%%                                                                            %%
%%****************************************************************************%%

% Begin of file SectionII.tex

\begin{section}{The Model Checker \MCMASSLK}\label{sec:mcmasslk}

\textbf{State Labelling Algorithm.}
The model checking algorithm for \SLK extends the corresponding ones for
temporal logic in two ways.
Firstly, it takes as input not only a formula, but also a binding which assigns
agents to variables.
Secondly, it does not merely return sets of states, but sets of pairs $\Tuple{g,
\asgFun}$ consisting of a state $g$ and an assignment of variables to strategies
$\asgFun$.
A pair $\Tuple{g, \asgFun} \in \ExtSet$ is called an \emph{extended state};
intuitively, $\asgFun$ represents a strategy assignment under which the formula
holds at state $g$.

		%% \begin{enumerate}
		%% 	\item
		%% 		It has a \emph{binding} $b \in \BndSet$ of agents to variables as
		%% 		the second argument.
		%% 	\item
		%% 		It calculates a set of \emph{extended states}.
		%% 		A extended state  $\Tuple{g, v}\in \ExtSet$ is a pair of \emph{(i)}
		%% a
		%% 		state $g \in G$, and \emph{(ii)} an assignment of variables to
		%% 		strategies $v \in \mathit{Asg}$ subject to which the formula holds in
		%% 		that state.
		%% \end{enumerate}

Given an \SLK formula $\varphi$ and a binding $b \in \BndSet \defeq \AgnSet \to
\VarSet$, the model checking algorithm $\SatFun\!: \mathit{\SLK} \times \BndSet
\rightarrow 2^{\ExtSet}$, returning a set of extended states, is defined as
follows, where $a \in \AgnSet$ is an agent, $\ASet \subseteq \AgnSet$ a set of
agents, and $x \in \VarSet$ a variable:
\begin{itemize}
% \item
% $\SatFun(\top, b) \defeq \ExtSet$.
\item
$\SatFun(p, b) \defeq \set{\Tuple{g, \asgFun} }{ g \in h(p) \wedge \asgFun \in
\AsgSet}$, with $p \in \APSet$;
\item
$\SatFun(\neg\varphi, b) \defeq \mathsf{neg}\!\left(\SatFun(\varphi, b)\right)$;
\item
$\SatFun(\varphi_1\wedge\varphi_2, b) \defeq \SatFun(\varphi_1, b) \cap
\SatFun(\varphi_2, b)$;
% \item
% $\SatFun(\varphi_1\vee\varphi_2, b) \defeq \SatFun(\varphi_1, b) \cup
% \SatFun(\varphi_2, b)$.
\item
$\SatFun(\B{a}{x}\varphi, \allowbreak b) \defeq \SatFun(\varphi, b[a \mapsto
x])$;
\item
$\SatFun(\EExs{x}\varphi, b) \defeq \set{\Tuple{g, \asgFun} }{ \exists \fFun \in
\StrSet[\mathsf{sharing}(\varphi, x)] . \Tuple{g, \asgFun\!\left[x \mapsto
\fFun\right]} \in \SatFun(\varphi, b)}$;
% \item
% $\SatFun(\AAll{x}\varphi, b) \defeq \set{\Tuple{g, v} }{\forall f \in
% \mathit{Str}_{\mathsf{sharing}(\varphi, x)} .  \Tuple{g, v[x \mapsto f]} \in
% \SatFun(\varphi, b)}$.
\item
$\SatFun(\X{}\varphi, b) \defeq \mathsf{pre}(\SatFun(\varphi, b), b)$;
\item
$\SatFun(\varphi_1\U\varphi_2, b) \defeq
\mathrm{lfp}_{X}\!\left[\SatFun(\varphi_2, b) \cup \left(\SatFun(\varphi_1, b)
\cap \mathsf{pre}(X, b)\right)\right]$;
\item
% $\SatFun(\K[a]\varphi, b) \defeq G \setminus \!\left(\set{\Tuple{g, \asgFun}
% }{ \exists \Tuple{g', \asgFun'} \in \SatFun(\neg\varphi, \varnothing). g'
% \sim_a g}\right)$.
$\SatFun(\K[a]\varphi, b) \defeq \mathsf{neg}\!\left(\set{\Tuple{g, \asgFun} }{
\exists \Tuple{g', \asgFun'} \in \SatFun(\neg\varphi, \varnothing). g' \sim_a
g}\right)$;
\item
$\SatFun(\D[\ASet]\varphi, b) \defeq \mathsf{neg}\!\left(\set{\Tuple{g, \asgFun}
}{ \exists \Tuple{g', \asgFun'} \in \SatFun(\neg\varphi, \varnothing). g'
\sim^{\D}_{\ASet} g}\right)$;
\item
$\SatFun(\C[\ASet]\varphi, b) \defeq \mathsf{neg}\!\left(\set{\Tuple{g, \asgFun}
}{ \exists \Tuple{g', \asgFun'} \in \SatFun(\neg\varphi, \varnothing). g'
\sim^{\C}_{\ASet} g}\right)$.
\end{itemize}
Above we use $h(p)$ to denote the set of global states where atom $p$ is true;
$\mathsf{pre}(C, b)$ is the set of extended states that temporally precede $C$
subject to a binding $b$;
%
%% $G$ is the set of reachable states for $\IName$,
%
$\mathsf{neg}(C)$ stands for the set of extended states $\Tuple{g, \asgFun}$
such that for each extended state $\Tuple{g, \asgFun'} \in C$, there is some
variable $x \in \dom{\chi} \cap \dom{\chi'}$, such that the strategies
$\asgFun(x)$ and $\asgFun'(x)$ disagree on the action to be carried out in some
global state $g' \in \dom{\chi(x)} \cap \dom{\chi'(x)}$ (\ie, $\chi(x)(g') \neq
\chi'(x)(g')$); $\StrSet[\mathsf{sharing}(\varphi, x)]$ is the set of strategies
shared by the agents bound to the variable $x$ in the formula $\varphi$;
finally, $\sim_a$, $\sim^{\D}_{\ASet}$, and $\sim^{\C}_{\ASet}$ represent the
individual, distributed, and common epistemic accessibility relations for agent
$\aElm$ and agents $\ASet$ defined on the respective notions of equality of
agents' local states.
The set of global states of an interpreted system $\mathcal{I}$ satisfying a
given formula $\varphi \in \mathit{\SLK}$ is calculated from the algorithm above
by computing $\left\|\varphi\right\|_{\mathcal{I}} \defeq \set{ g \in G }{
\Tuple{g, \varnothing} \in \SatFun(\varphi, \varnothing)}$.

\textbf{BDD Translation.}
Given an interpreted system $\mathcal{I}$ and an \SLK formula $\varphi$, we now
summarise the steps required to implement the labelling algorithm above using
OBDDs~\cite{Bry86}.
We represent global states and joint actions as Boolean vectors $\overline{v}$
and $\overline{w}$, respectively~\cite{RL07}.
Similarly, an assignment $\asgFun$ is represented as a Boolean vector
$\overline{u}$ with $K = \sum_{x \in \VarSet} \allowbreak \sum_{S \in G
/\sim^{\C{}}_{\mathsf{sharing}\left(\varphi, x\right)}} \left\lceil
\log_2\left|\bigcap_{g \in S}\bigcap_{a \in \mathsf{sharing}\left(\varphi,
x\right)}P_a(l_a(g))\right|\right\rceil$ Boolean variables.
%% where
%% $\sim_{\ASet}^{\C{}}$ denotes the common knowledge relation for agents
%% $\ASet \subseteq \AgnSet$.
Intuitively, for each variable $x \in \VarSet$ and set of shared local states $S
\in G /\! \sim^{\C{}}_{\mathsf{sharing} \left(\varphi, x\right)}$, we store
which action should be carried out.
%AL: Really not here.
%% This demonstrates the large
%% number of boolean variables needed to encode extended states, which
%% constitutes a major bottleneck in SLK verification.
An extended state $\Tuple{g, \asgFun} \in \ExtSet$ is then represented as a
conjunction of the variables in $\overline{v}_g$ and $\overline{u}_{\asgFun}$.
%AL: Likewise
%% Note that finding a more compact BDD representation is highly unlikely due to
%% the need to represent sets of extended states (\ie, sets of functions).

Given a binding $b \in \BndSet$, we encode the protocol
$P(\overline{v}, \overline{w})$, the evolution function $t(\overline{v},
\overline{w}, \overline{v'})$, and the strategy restrictions $S^b(\overline{v},
\overline{w}, \overline{u})$, as in~\cite{LR06a}.
The temporal transition is encoded as $R^b_t(\overline{v}, \overline{v'},
\overline{u}) = \bigvee_{\overline{w} \in \mathit{Act}} t(\overline{v},
\overline{w}, \overline{v'}) \wedge P(\overline{v}, \overline{w}) \wedge
S^b(\overline{v}, \overline{w}, \overline{u})$.
Observe that we quantify over actions, encoded as $\overline{w}$, as
in~\cite{LR06a}, but we store the variable assignment in the extra parameter
$\overline{u}$.
Quantification over the variable assignment is performed when a strategy
quantifier is encountered.

Given this, the algorithm $\SatFun\!\left(\cdot, \cdot\right)$ is translated
into operations on BDDs representing the encoded sets of extended states.

\textbf{Implementation and Usage.}
The model checker \MCMASSLK~\cite{MCMASSLK} contains an implementation of the
procedure described previously.
To do this, we took \MCMAS as baseline~\cite{LQR09}.
\MCMAS is an open-source model checker for the verification of multi-agent
systems against \ATL and epistemic operators.
We used \MCMAS to parse input and used some of its existing libraries for
handling counter-examples, which were extended to handle \SLK modalities.

\MCMASSLK takes as input a system description given in the form of an ISPL
file~\cite{LQR09} providing the agents in the system, their possible local
states, their protocols, and their evolution functions.
Upon providing \SLK specifications, the checker calculates the set of reachable
extended states, encoded as OBDDs, and computes the results by means of the
labelling algorithm described previously.
If the formula is not satisfied, a counterexample is provided in the form of
strategies for the universally quantified variables.

\end{section}

% End of file SectionII.tex

%%****************************************************************************%%
%%                                                                            %%
%% A Model Checker for the Verification of Strategy Logic Specifications      %%
%%                                                                            %%
%% SectionIII.tex                                                             %%
%%                                                                            %%
%% Revision 1                                                                 %%
%%                                                                            %%
%% Copyright (C) 2014, Petr Cermak, Alessio Lomuscio, Fabio Mogavero, and     %%
%%                     Aniello Murano.                                        %%
%% All rights reserved.                                                       %%
%%                                                                            %%
%%****************************************************************************%%

% Begin of file SectionIII.tex

\begin{section}{Experimental Results and Conclusions}
	\label{sec:exprescon}

\textbf{Evaluation.}
To evaluate the proposed approach, we present the experimental results obtained
on the dining cryptographers protocol~\cite{Cha88,LQR09} and a variant of the
cake-cutting problem~\cite{ES84}.
The experiments were run on an Intel Core i7-2600 CPU 3.40GHz machine with 8GB
RAM running Linux kernel version 3.8.0-34-generic.
Table~\ref{tab:dining-cryptographers} reports the results obtained when
verifying the dining cryptographers protocol against the specifications
$\phi_{\CTLK} \defeq \A\G \psi$ and $\phi_{\SLK} \defeq \wp \G \psi$, with
$\AAll{\xElm} \varphi \defeq \neg \EExs{\xElm} \neg \varphi$, where:
\begin{align*}
\psi &\defeq \left(\mathrm{odd} \!\wedge\! \neg\mathrm{paid}_1\right)
\rightarrow \left(\K[\mathrm{c}_1]\! \left(\mathrm{paid}_2 \vee \!\cdots\! \vee
\mathrm{paid}_n\right)\right) \wedge \left(\neg \K[\mathrm{c}_1] \mathrm{paid}_2
\wedge \!\cdots\! \wedge \neg \K[\mathrm{c}_1] \mathrm{paid}_n\right) \\
\wp &\defeq {\AAll{x_1} \cdots \AAll{x_n} \AAll{x_{\mathrm{env}}}}
\B{\mathrm{c}_1}{x_1}
\, \cdots \B{\mathrm{c}_n}{x_n} \B{\mathrm{Environment}}{x_{\mathrm{env}}}
\end{align*}

%% Table ~\ref{tab:dining-cryptographers} shows the results for different
%% numbers of cryptographers.
%%  It is difficult to compare the performance
%% of \SLK formula checking because there are currently no other existing
%% tools supporting \SLK.  Even more, there are no ones neither for the
%% original \SL nor for \ATLS.

\vspace{-2em}
\begin{table}
\centering
\caption{Verification results for the dining cryptographers protocol.}
\begin{tabular}{|c|c|c|c|c|c|}
\hline
$n$ \textbf{crypts} & \textbf{possible states} & \textbf{reachable states}
& \textbf{reachability} (s) & \textbf{CTLK} (s) & \textbf{SLK} (s) \\ \hline
$10$ & $3.80 \times 10^{14}$ & $45056$ & $4.41$ & $0.30$ & $2.11$ \\ \hline
$11$ & $9.13 \times 10^{15}$ & $98304$ & $1.79$ & $0.04$ & $5.51$ \\ \hline
$12$ & $2.19 \times 10^{17}$ & $212992$ & $2.43$ & $0.02$ & $11.78$ \\ \hline
$13$ & $5.26 \times 10^{18}$ & $458752$ & $2.17$ & $0.11$ & $32.41$ \\ \hline
$14$ & $1.26 \times 10^{20}$ & $983040$ & $2.08$ & $0.09$ & $85.29$ \\ \hline
$15$ & $3.03 \times 10^{21}$ & $2.10 \times 10^6$ & $22.67$ & $0.33$ & $171.61$
\\ \hline
$16$ & $7.27 \times 10^{22}$ & $4.46 \times 10^6$ & $7.13$ & $0.09$ & $451.41$
\\ \hline
$17$ & $1.74 \times 10^{24}$ & $9.44 \times 10^6$ & $9.77$ & $0.13$ & $768.34$
\\ \hline
\end{tabular}
\label{tab:dining-cryptographers}
\end{table}
\vspace{-1em}

$\phi_{\CTLK}$ is the usual epistemic specification for the
protocol~\cite{MS04,LQR09} and $\phi_{\SLK}$ is its natural extension where
strategies are quantified.
The results show that the checker can verify reasonably large state spaces.
The performance depends on the number of Boolean variables required to represent
the extended states.
In the case of \SLK specifications, the number of Boolean variables is
proportional to the number of strategies (here equal to the number of agents).
The last two columns of Table~\ref{tab:dining-cryptographers} show that the
tool's performance drops considerably faster when verifying \SLK formulas
compared to \CTLK ones.
This is because \CTLK requires no strategy assignments and extended states
collapse to plain states.
In contrast, the performance for \CTLK is dominated by the computation of the
reachable state space.

We now evaluate \MCMASSLK  with respect to strategy synthesis and specifications
expressing Nash equilibria.
Specifically, we consider a variation of the model for the classic cake-cutting
problem~\cite{ES84} in which a set of \emph{$n$ agents} take turns to slice a
cake of size $d$ and the \emph{environment} responds by trying to ensure the
cake is divided fairly.
We assume that at each even round the agents concurrently choose how to divide
the cake; at each odd round the environment decides how to cut the cake and how
to assign each of the pieces to a subset of the agents.
Therefore, the problem of cutting a cake of size $d$ between $n$ agents is
suitably divided into several simpler problems in which pieces of size $d' < d$
have to be split between $n' < n$ agents.
The multi-player game terminates once each agent receives a slice.

The model uses as atomic propositions pairs
$\left\langle i, c \right\rangle \in \numcc{1}{n} \times \numcc{1}{d}$
indicating that agent $i$ gets a piece of cake of size $c$.
The existence of a protocol for the cake-cutting problem is given by the
following \SL specification $\varphi$:
%\vspace{-0.5em}
\[
\varphi \defeq \EExs{\xElm} (\varphi_{F} \wedge \varphi_{S}), \text{where}
\]
%\vspace{-2.5em}
\begin{itemize}
\item
$\varphi_{F} \defeq \AAll{\yElm[1]} \ldots \AAll{\yElm[n]} (\psi_{\mathit{NE}}
\rightarrow \psi_{E})$ ensures that the protocol $\xElm$ is fair, \ie, all Nash
equilibria $(\yElm[1], \ldots, \yElm[n])$ of the agents guarantee equity of
the splitting;
\item
$\varphi_{S} \defeq \EExs{\yElm[1]} \ldots \EExs{\yElm[n]} \psi_{\mathit{NE}}$
ensures
that the protocol has a solution, \ie, there is at least one Nash equilibrium;
\item
$\psi_{\mathit{NE}} \defeq \bigwedge_{i = 1}^{n} (\bigwedge_{v = 1}^{d}
(\EExs{\zElm}
\bndElm[i] \pElm[i](v)) \rightarrow (\bigvee_{c = v}^{d} \bndElm \pElm[i](c)))$
ensures that if agent $i$ has a strategy $\zElm$ allowing him to get from the
environment a slice of size $v$ once the strategies of the other agents are
fixed, he is already able to obtain a slice of size $c \geq v$ by means of his
original strategy $\yElm[i]$ (this can be ensured by taking $\bndElm \!\defeq\!
(\mathrm{Environment}, \xElm) (1, \yElm[1]) \ldots (n, \yElm[n])$, $\bndElm[i]
\!\defeq\! (\mathrm{Environment}, \xElm) (1, \yElm[1]) \cdots (i, \zElm) \cdots
\allowbreak (n, \yElm[n])$, and $\pElm[i](c) \!\defeq\! \F
\left\langle i, c \right\rangle$);
\item
$\psi_{E} \defeq \bndElm \bigwedge_{i = 1}^{n} \pElm[i](\floor{d / n})$ ensures
that each agent $i$ is able to obtain a piece of size $\floor{d / n}$ ($\bndElm$
and $\pElm[i]$ are the same as in the item above).
\end{itemize}
We were able to verify the formula $\varphi$ defined above on a system with $n =
2$ agents and a cake of size $d = 2$.
Moreover, we automatically synthesised a strategy $x$ for the environment
(see~\cite{MCMASSLK} for more details).
We were not able to verify larger examples; for example with $n=2, d=3$, there
are $29$ reachable states; the encoding required $105$ Boolean variables (most
of them represent the assignments in the sets of extended states), and the
intermediate BDDs were found to be in the order of $10^9$ nodes.
This should not be surprising given the theoretical difficulty of the
cake-cutting problem.
Moreover, we are synthesising the entire protocol and not just the agents'
optimal behaviour.
%% Even though this is obviously negative, we are not aware of
%% other comparable results in the literature.
%AL: This is repeated below - I deleted it.

\textbf{Conclusions.}
In this paper we presented \MCMASSLK, a novel symbolic model checker for the
verification of systems against specifications given in \SLK.
A notable feature of the approach is that it allows for the automatic
verification of sophisticated game concepts such as various forms of equilibria,
including Nash equilibria.
Since \MCMASSLK also supports epistemic modalities, this further enables us to
express specifications concerning individual and group knowledge of cooperation
properties; these are commonly employed when reasoning about multi-agent
systems.
Other tools supporting epistemic or plain \ATL specifications
exist~\cite{AHMQRT98,GM04,KNNPPSWZ07,LQR09}.
In our experiments we found that the performance of \MCMASSLK on the \ATL and
\CTLK fragments was comparable to that of \MCMAS, one of the leading checkers
for multi-agent systems.
This is because we adopted an approach in which the colouring with strategies is
specification-dependent and is only performed after the set of reachable states
is computed.

As described, a further notable feature of \MCMASSLK is the ability to
synthesise behaviours for multi-player games, thereby going beyond the
classical setting of two-player games.

We found that the main impediment to better performance of the tool is
the size of the BDDs required to encode sets of extended
states. Future efforts will be devoted to mitigate this problem as
well as to support other fragments of SL.

%% In the future we intend to parallelise some of the routines developed in order
%% to speed-up the verification time of \SLK specifications.

\end{section}

\textbf{Acknowledgements.}
This research was partly funded by the EPSRC (grant EP/I00520X), the
Regione Campania (Embedded System Cup project B25B09090 100007), the
EU (FP7 project 600958-SHERPA), and the MIUR (ORCHESTRA
project). Aniello Murano acknowledges support from the Department of
Computing at Imperial College London for his research visit in July
2013.

%\end{section}

% End of file Acknowledgments.tex

\bibliographystyle{plain}
\bibliography{References}

\end{document}